\documentclass[10pt]{article}

\usepackage{amsmath}
\usepackage{amssymb}

\usepackage{subfig}
\usepackage{graphicx}
\usepackage{algorithmic}
\usepackage{url}
\usepackage{tabularx}
\usepackage{times}
\usepackage[iso-8859-15]{inputenx}
\usepackage{textcomp}
\usepackage{balance}
\usepackage{tabu}
\usepackage{dblfloatfix}
\usepackage{enumitem}
\usepackage{verbatim}
\usepackage{cite}
\usepackage[ruled]{algorithm2e}
\usepackage{color} 
\usepackage{nameref}
\usepackage{xcolor}

\definecolor{darkgreen}{HTML}{000000}
\newcommand{\green}[1]{{\color{darkgreen}#1}}

\usepackage[labelfont=bf,labelsep=period,justification=raggedright]{caption}

\makeatletter
\renewcommand{\@biblabel}[1]{\quad#1.}
\makeatother

\date{}

\SetAlFnt{\small}
\SetAlCapFnt{\small}
\SetAlCapNameFnt{\small}
\SetAlCapHSkip{0pt}
\IncMargin{-\parindent}

\begin{document}

\begin{flushleft}
{\Large
\textbf{Detecting Memory and Structure in Human Navigation Patterns Using Markov Chain Models of Varying Order}
}
\\
Philipp Singer$^{1,\ast}$, 
Denis Helic$^{2}$, 
Behnam Taraghi$^{3}$,
Markus Strohmaier$^{1,4}$
\\
\bf{1} GESIS - Leibniz Institute for the Social Sciences, Cologne, Germany
\\
\bf{2} Technical University of Graz, Knowledge Technologies Institute, Graz, Austria
\\
\bf{3} Technical University of Graz, Institute for Information Systems and Computer Media, Graz, Austria
\\
\bf{4} University Koblenz-Landau, Institute for Web Science and Technologies, Koblenz, Germany
\\
$\ast$ E-mail: philipp.singer@gesis.org
\end{flushleft}

\section*{Abstract}
One of the most frequently used models for understanding human navigation on the
Web is the Markov chain model, where Web pages are represented as states and hyperlinks
as probabilities of navigating from one page to another.
Predominantly, human navigation on the Web has been thought to satisfy the memoryless
Markov property stating that the next page a user visits only depends on her
current page and not on previously visited ones.
This idea has found its way in numerous applications such as Google's PageRank
algorithm and others. Recently, new studies suggested that human navigation may better be modeled using higher order Markov chain models, i.e., the next page depends on a longer history of
past clicks.
Yet, this finding is preliminary and does not account for the higher complexity of higher order Markov chain models which is why
the memoryless model is still widely used. 
In this work we thoroughly present a diverse array of advanced inference methods for determining the appropriate Markov chain order. We highlight strengths and weaknesses of each method and apply them for investigating memory and structure of human navigation on the Web.
Our experiments reveal that the complexity of higher order models grows faster
than their utility, and thus we confirm that the memoryless model represents a
quite practical model for human navigation on a page level. However, when we
expand our analysis to a topical level, where we abstract away from specific
page transitions to transitions between topics, we find that the memoryless
assumption is violated and specific regularities can be observed.
We report results from experiments with two types of navigational datasets
(goal-oriented vs.
free form) and observe interesting structural differences that make a strong
argument for more contextual studies of human navigation in future work.

\section*{Introduction}
\label{sec:intro}

Navigation represents a fundamental activity for users on the Web. Modeling
this activity, i.e., understanding how predictable human navigation
is and whether regularities can be detected has been of interest to researchers for
nearly two decades -- an example of early work would be work by Catledge and Pitkow~\cite{catledge}.
Another example would be \cite{xing}, who focused on trying to understand preferred user navigation patterns in
order to reveal users' interests or preferences.
Not only has our community been interested in gaining deeper insights into human
behavior during navigation, but also in understanding how models of human
navigation can improve user interfaces or information network structures
\cite{borges1999}.
Further work has focused on understanding whether models of human navigation can
help to predict user clicks in order to prefetch Web sites (e.g.,
\cite{bestavros}) or enhance a site's interface or structure (e.g.,
\cite{perkowitz}).
More recently, such models have also been deployed in the field of
recommender systems (e.g., \cite{rendle}).



However, models of human navigation can only be useful to the extent human
navigation itself exhibits regularities that can be exploited. An early study on
user navigation in the Web by Huberman et al.~\cite{huberman}, for example,
already identified interesting regularities in the distributions of user page
visits on a Web site. 
More recently, Wang and Huberman~\cite{wang2012} confirmed these observations
and Song et al. \cite{song} argued that the regularities in human activities
might be based on the inherent regularities of human behavior in general.

The most prominent model for describing human navigation on the Web is the
Markov chain model (e.g., \cite{pirolli}), where Web pages are represented as
states and hyperlinks as probabilities of navigating from one page to another.
Predominantly, the Markov chain model has been memoryless in a wide range of
works (e.g., Google's PageRank \cite{brin}) indicating that the next state only depends on the
current state of a user's Web trail. Recently, a study \cite{chierichetti}
suggested that human navigation might be better modeled with memory -- i.e., the next
page depends on a longer history of past clicks. However, this finding is preliminary and does not account for the higher complexity of higher order Markov chain models which is why
the memoryless model is still widely used.

\begin{figure}[t!] \centering
\includegraphics[width=0.85\textwidth]{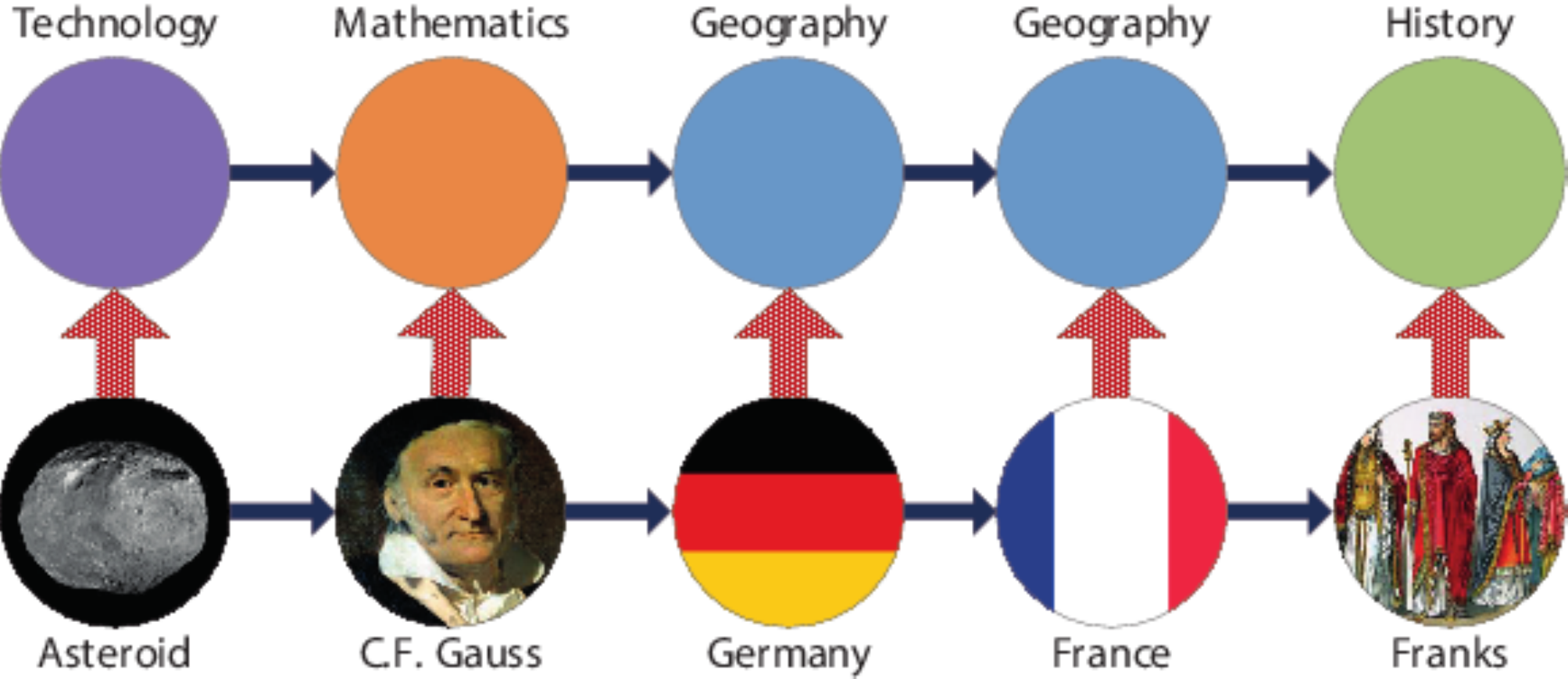}
\caption{\textbf{Example of a navigation sequence in the WikiGame dataset.}
Bottom row of nodes: A user navigates a series of Wikipedia articles,
which can be represented as a sequence of Web pages. Top row of nodes:
Each Wikipedia article can be mapped to a corresponding topic
through Wikipedia's system of categories. This results in a sequence of topics.}
\label{fig:pathexample}
\end{figure}

\smallskip \noindent {\bf Research questions.} In this paper, we are interested
in shedding a deeper light on regularities in human navigation on the World Wide
Web by studying memory and structure in human navigation patterns.
We start by investigating memory of human navigational paths over Web sites
by determining the order of corresponding Markov chains. We are specifically
interested in detecting if the benefit of a larger memory (or higher order
Markov chain) can compensate for the higher complexity of the model. In order to
understand whether and to what extent human navigation exhibits memory on
a topical level, we abstract away from specific page transitions and
study memory effects on a topical level by representing click streams as
sequences of topics\footnote{Note that the terms "`topic"' and "`category"' should be seen as synonyms throughout this work."'} (cf.
Figure \ref{fig:pathexample}). This enables us to (i) move up from the page to
topical level and (ii) significantly reduce the complexity of higher order
models and therefore (iii) gain deeper insights into memory and structure of
human navigational patterns. Finally, we discuss our findings and demonstrate
interesting differences between human navigation in free browsing vs. more
goal-oriented settings.

\smallskip \noindent {\bf Methods and Materials.} We study memory and
structure in human navigation patterns on three similarly structured
datasets: WikiGame (a navigation dataset with known navigation goals),
Wikispeedia (another goal-oriented navigation dataset) and MSNBC (a free navigation
dataset). 
\green{
For analyzing memory, we use Markov chains to model human behavior and
analyze the appropriate Markov chain order -- i.e., we investigate
whether human navigation is memoryless or not.
For model selection -- i.e., the process of finding the most appropriate Markov chain order -- we resort to a highly diverse array of methods stemming from distinct statistical schools:
(i) likelihood \cite{stigler2002statistics,tong1975}, (ii) Bayesian \cite{Strelioff} and (iii) information-theoretic methods \cite{akaike,katz,murphy,schwarz,tong1975}. We supplement these with a (iv) cross validation approach for a prediction task \cite{murphy}. We thoroughly elaborate each method, put them into relation to each other and also highlight strengths and weaknesses of each.
Such
detailed derivation of model parameters and the model comparison is, for example, missing in previous work
\cite{chierichetti}, which prevents us from drawing definite conclusions.
We apply these methods to our human navigational data in order to get an exhaustive picture about memory in human navigation. 
Finally, we identify structural aspects by analyzing
transition matrices produced by our Markov chain analyses.
}

\smallskip
\noindent {\bf Contributions.} The main contributions of this work are three-fold:
\begin{itemize}
\item First, we deploy four different, yet complementary, approaches for order selection of Markov chain models
(likelihood, Bayesian, information-theoretic and cross validation methods) and elaborate their strengths and weaknesses. \green{Hence, our work extends existing studies that
model human navigation on the Web using Markov chain models \cite{chierichetti}.}
By applying these methods on navigational Web data,
our work presents -- to the best of our knowledge -- the most comprehensive and
systematic evaluation of Markov model orders for human navigational sequences on the Web to date.
Furthermore, we make our methods in the form of an open source framework available online\footnote{\url{https://github.com/psinger/PathTools}} to aid future work \cite{github}.

\item Our empirical results confirm what we inferred from theory: It is
difficult to make plausible statements about the appropriate Markov chain order
having insufficient data but a vast amount of states, which is a common
situation for Web page navigational paths. All evaluation approaches would favor
a zero or first order because the number of parameters grows exponentially with the chain order and the available data is
too sparse for proper parameter inferences.
Thus, we show further evidence that the memoryless model seems to be a quite
practical and legitimate model for human navigation on a page level.

\item By abstracting away from the page level to a topical level, the results are different. 
By representing all datasets as navigational sequences of topics that describe
underlying Web pages (cf.
Figure \ref{fig:pathexample}), we find evidence that topical navigation of humans is not
memoryless at all.  On three rather different datasets of navigation -- free navigation (MSNBC) and
goal-oriented navigation (WikiGame and Wikispeedia) -- we find mostly consistent
memory regularities on a topical level: In all cases, Markov chain models of
order two (respectively three) best explain the observed navigational sequences.
We analyze the structure of such navigation, identify strategies and the
most salient common sequences of human navigational patterns and provide visual
depictions. 
Amongst other structural differences between goal-oriented and free form navigational patterns, users seem to stay in the same topic more frequently for our free form navigational dataset (MSNBC) compared to both of the goal oriented
datasets (Wikigame and Wikispeedia). 
Our analysis thereby provides new insights into
the memory and structure that users employ when navigating the Web that can e.g., be useful to improve recommendation algorithms, web site design or faceted browsing.
\end{itemize}

The paper is structured as follows: In the section entitled "\nameref{sec:related}" we review the
state-of-the-art in this domain. Next, we present our methodology and
experimental setup in the sections called "\nameref{sec:methodology}" and
"\nameref{sec:experimental}".
We present and discuss our results in the section named~"\nameref{sec:results}". In the section called "\nameref{subsec:discussion} we provide a final discussion and the section
called "\nameref{sec:conclusions}" concludes our paper.

\section*{Related Work}
\label{sec:related}


In the late 1990s, the analysis of user navigational behavior on
the Web became an important and wide-spread research topic.
Prominent examples are models by Huberman and Adamic~\cite{huberman2} that determine how
users choose new sites while navigating, or the work by Huberman et
al.~\cite{huberman} who have shown that strong regularities in human navigation behavior exist and that, for example, the length of navigational paths on the Web is
distributed as an inverse Gaussian distribution.
These first models of human navigation on the Web set a standard modeling framework for future research - the majority
of navigation models have been stochastic henceforth.
Common stochastic models of human navigation are Markov chains. For example,
the Random Surfer model in Google's PageRank algorithm can be seen as a special case of a Markov chain \cite{brin}.
Some further examples of the application of Markov chains as models of Web navigation can be found in
\cite{borges,deshpande,lempel,pirolli,sen,anderson,cadez,zukerman,pitkow}. 

 In a Markov chain, Web pages are represented as states and links between the
 pages are modeled as probabilistic transitions between the states.
 The dynamics
 of a user's navigation session, in which she visits a number of pages by
 following the links between them, can thus be represented as a sequence of
 states.
Specific configurations of model parameters -- such as transition
probabilities or model orders -- have been used to reflect different assumptions
about navigation behavior. One of the most influential assumptions in this field
to date is the so-called Markovian property, which postulates that the next
page that a user visits depends only on her current page, and not on any other
page leading to the current one.
This assumption is adopted in a number of prevalent models of human navigation
in information networks, for example also in the Random Surfer model
\cite{brin}. However, this property is neglecting the observations stated above that human
navigation exhibits strong regularities which hints towards longer memory
patterns in human navigation. We argue, that the more
consistency human navigation in information networks displays the higher the
appropriate Markov chain order should be.

\textit{The Markovian assumption might be wrong:}
The principle that human navigation might exhibit longer memory patterns than
the first order Markov chain captures has been investigated in the past (see
e.g., \cite{borges1999,pirolli} or \cite{rosvall2013networks} for a more general approach of looking at memory in network flows). However, higher order Markov chains have been
often disputed for modeling human navigation because the gain of a higher order model did not compensate for
the additional complexity introduced by the model \cite{pirolli}. Therefore, it
was a common practice to focus on a first order model since it was a reasonable
but extremely simple approximation of user navigation behavior (e.g.,
\cite{cadez,sarukkai,sen,zukerman}).

The discussion about the appropriate Markov chain order was just recently picked
up again by Chierichetti et al.~\cite{chierichetti}. While the authors' results
again show indicators that users on the World Wide Web are not Markovian, the
study does not account for the higher complexity of such models and the possible
lack of statistically significant gains of these models.
Technically, the authors analyzed Markov chain models of different orders by
measuring the likelihood of real navigational sequences given a particular
model. In the next step, the authors compared the models by their likelihoods
and found that the Markovian assumption does not hold for their given data and,
thus, higher order Markov chain models seem to be more appropriate. As a result,
the authors argue that users on the World Wide Web are not Markovian.
However, their results come with certain limitations, such as the fact that
choosing the model with the highest likelihood is biased towards models with
more parameters.
\green{Because lower order models are always nested within  higher order models and as higher order Markov chains have exponentially more parameters than lower
order models (potential overfitting), they are always a better fit for the data \cite{murphy}.} Thus, higher order
models are naturally favored by their improvements in likelihoods. A
more comprehensive view on this issue shows that there exists a broad range of established model
comparison techniques that also take into the account the complexity of a model
in question \cite{akaike,bartlett,gates1976,katz,schwarz,Strelioff,tong1975}.

Moreover, the principle objects of interest in the majority of the past studies
are transitions between Web pages. Only a few studies \cite{cadez,kumar2010,west} investigate navigation as transitions between Web page
features, such as the content or context of those Web pages. 


\section*{Methods}
\label{sec:methodology}

\green{
In the following, we briefly introduce Markov chains before discussing an
expanded set of methods for order selection, including \textit{likelihood}, \textit{Bayesian}, \textit{information-theoretic} and \textit{cross
validation} model selection techniques. 
}

\subsection*{Markov Chains}
\label{subsec:markovchains}

Formally, a discrete (time and space) finite Markov chain is a stochastic process which amounts to a sequence of random variables $X_1, X_2, ..., X_n$. For a Markov
chain of the first order, i.e., for a chain that satisfies the memoryless Markov property
the following holds:
\begin{eqnarray}
\nonumber P(X_{n+1} = x_{n+1} | X_1 = x_1, X_2 = x_2, ..., X_n = x_n) & = \\
 P(X_{n+1} = x_{n+1} | X_n = x_n)
\end{eqnarray}

\green{
This classic first order Markov chain model is usually also called a \emph{memoryless model} as we only use the current information for deriving the future and do not look into the past.}
For all our models we assume \textit{time-homogeneity} -- the probabilities do not change as a function of time.
To simplify the notation we denote data as a sequence $D=(x_1, x_2, ..., x_n)$ with states from a finite set $S$. With this
simplified notation we write the Markov property as:
\begin{equation}
 p(x_{n+1}|x_1, x_2, ..., x_n)=p(x_{n+1}|x_n)
\end{equation}

\green{
As we are also interested in higher order Markov chain models in this article -- i.e., memory models -- we now also define a Markov chain for an arbitrary order $k$ with $k \in \mathbb{N}$ -- or a chain with memory $k$.
}
In a Markov chain of $k$-th order the probability of the next state depends on $k$ previous states. 
Formally, we write:
\begin{equation}
 p(x_{n+1}|x_1, x_2, ..., x_n)=p(x_{n+1}|x_n, x_{n-1}, ..., x_{n-k+1})
\end{equation}

Markov chains of a higher order can be converted into Markov chains of order one in a straightforward manner -- the set of states for a higher order Markov chain includes all
sequences of length $k$ (resulting in a state set of size $|S|^k|S|$). The transition probabilities are adjusted accordingly.

A Markov model is typically represented by a transition (stochastic) matrix
$P$ with elements $p_{ij}=p(x_j|x_i)$. Since $P$ is a stochastic matrix it holds that for all $i$:
\begin{equation}
 \sum_j p_{ij} = 1
\end{equation}

Please note, that for a Markov chain of order $k$ the current state $x_i$ can be
a compound state of length $k$ -- it is a sequence of past $k$ states. 
\green{Throughout this paper we use this simpler notation, but one should keep in mind that $x_i$ differs for distinct orders $k$.}

\green{
For the sake of completeness, we also allow $k$ to be zero. In such a \emph{zero order} Markov chain model the next state does not depend on any current or previous events, but simply can be seen as a \emph{weighted random selection} -- i.e., the probability of choosing a state is defined by how frequently it occurs in the navigational paths. This should serve as a baseline for our evaluations.}

Next, we want to estimate the vector $\theta$ of parameters
of a particular Markov chain that generated observed data $D$ as well as determine the appropriate Markov chain order. For a Markov chain the model parameters are the elements $p_{ij}$ of the
transition matrix $P$, i.e., $\theta = P$.


\subsection*{Model Selection}
\label{subsec:modelselection}

\green{
In this article our main goal is to determine the appropriate order of a Markov chain -- i.e., the
appropriate length of the memory. For doing so, we resort to well established statistical methods. As we want to provide a preferably complete array of methods for doing so, we present and apply methods from distinct statistical schools: (i) likelihood, (ii) Bayesian and (iii) information-theoretic methods\footnote{\green{
Note that no official classification of statistical schools is available; some may also argue that there are only the two competing schools of frequentists (which we do not explicitly discuss in this article) and Bayesians. The categorization used here is motivated by a short blog post (see \url{http://labstats.net/articles/overview.html}).
}
}.
We also supplement the methods coming from these three schools by providing a model selection technique usually known from machine learning: (iv) cross validation.
We provide an overall ample view of methods and discuss advantages and limitations of each in the following sections.
}


\subsubsection*{Likelihood Method}
\label{subsubsec:mle}

\green{
The term \emph{likelihood} was coined and popularized by R. A. Fisher in the 1920's (see e.g, \cite{stigler2002statistics} for a historic recap of the developments).
Likelihood can be seen as a central element of statistics and we will also see in the following sections that other methods also resort to the concept. 
}
The likelihood is a function of the parameters
$\theta$ and it equals to the probability of observing the data given specific
parameter values:
\begin{eqnarray}
\nonumber P(D|\theta) &=& p(x_n|x_{n-1})p(x_{n-1}|x_{n-2})...p(x_2|x_1)p(x_1)\\
 &=& p(x_1) \prod_i \prod_j p_{ij}^{n_{ij}},
\end{eqnarray}
where $n_{ij}$ is the number of transition from state $x_i$ to state $x_j$ in $D$.

Fisher also popularized the so-called \emph{maximum likelihood estimate (MLE)}
which has a very intuitive interpretation.
This is the estimation of the parameters $\theta$ -- i.e., transition probabilities -- that most
likely generated data $D$. 
Concretely, the maximum likelihood estimate $\hat{\theta}_{MLE}$ are the values of the
parameters $\theta$ that maximize the likelihood function, i.e.,
$\hat{\theta}_{MLE}=\arg\max_\theta P(D|\theta)$ (a thorough introduction to MLE can be found in \cite{royall1997statistical}).
%

 The maximum likelihood estimation for Markov chains is an example of an optimization
 problem under constraints. Such optimization problems are typically solved by
 applying Lagrange multipliers. To simplify the calculus we will work with the
 log-likelihood function $\mathcal{L(P(D|\theta))}=log P(D|\theta)$. Because the
 $log$ function is a monotonic function that preserves order, maximizing the
 log-likelihood is equivalent to maximizing the likelihood function. Thus, we
 have:
 \begin{eqnarray}
  \nonumber \mathcal{L(P(D|\theta))} &=& log \left(p(x_1) \prod_i \prod_j p_{ij}^{n_{ij}}\right) \\
  &=& log p(x_1) + \sum_i \sum_j n_{ij} log p_{ij}
 \end{eqnarray}

 Our constraints capture the fact that each transition matrix row sums to $1$:
 \begin{equation}
  \sum_j p_{ij} = 1
 \end{equation}

 We have $n$ rows and therefore we need $n$ Lagrange multipliers $\lambda_1, \lambda_2, ..., \lambda_n$. We can rewrite the constraints using Lagrange multipliers as:
 \begin{equation}
  \lambda_i\left(\sum_j p_{ij} - 1\right) = 0
 \end{equation}

 Now, the new objective function is:
 \begin{equation}
  f(\mathbf{\lambda, \theta}) = \mathcal{L}(P(D|\theta)) - \sum_i\lambda_i\left(\sum_j p_{ij} - 1\right)
 \end{equation}

 To maximize the objective function we set partial derivatives with respect to $\lambda_i$ to $0$, which gives back the original constraints.

 Further, we set partial derivatives
 with respect to $p_{ij}$ to $0$ and solve the equation system for $p_{ij}$. This gives:
 \begin{equation}
  p_{ij} = \frac{n_{ij}}{\sum_j n_{ij}}
 \end{equation}

Thus, the maximum likelihood estimate for a specific $p_{ij}$ is the number of
transitions from state $x_i$ to state $x_j$ divided by the total number of
transitions from state $x_i$ to any other state. For example, in a navigation
scenario the maximum likelihood estimate for a transition from page $A$ to page $B$ is the
number of clicks on a link leading to page $B$ from page $A$ divided by the
total number of clicks on page $A$.

\begin{figure}[t!] \centering
\includegraphics[width=0.7\columnwidth]{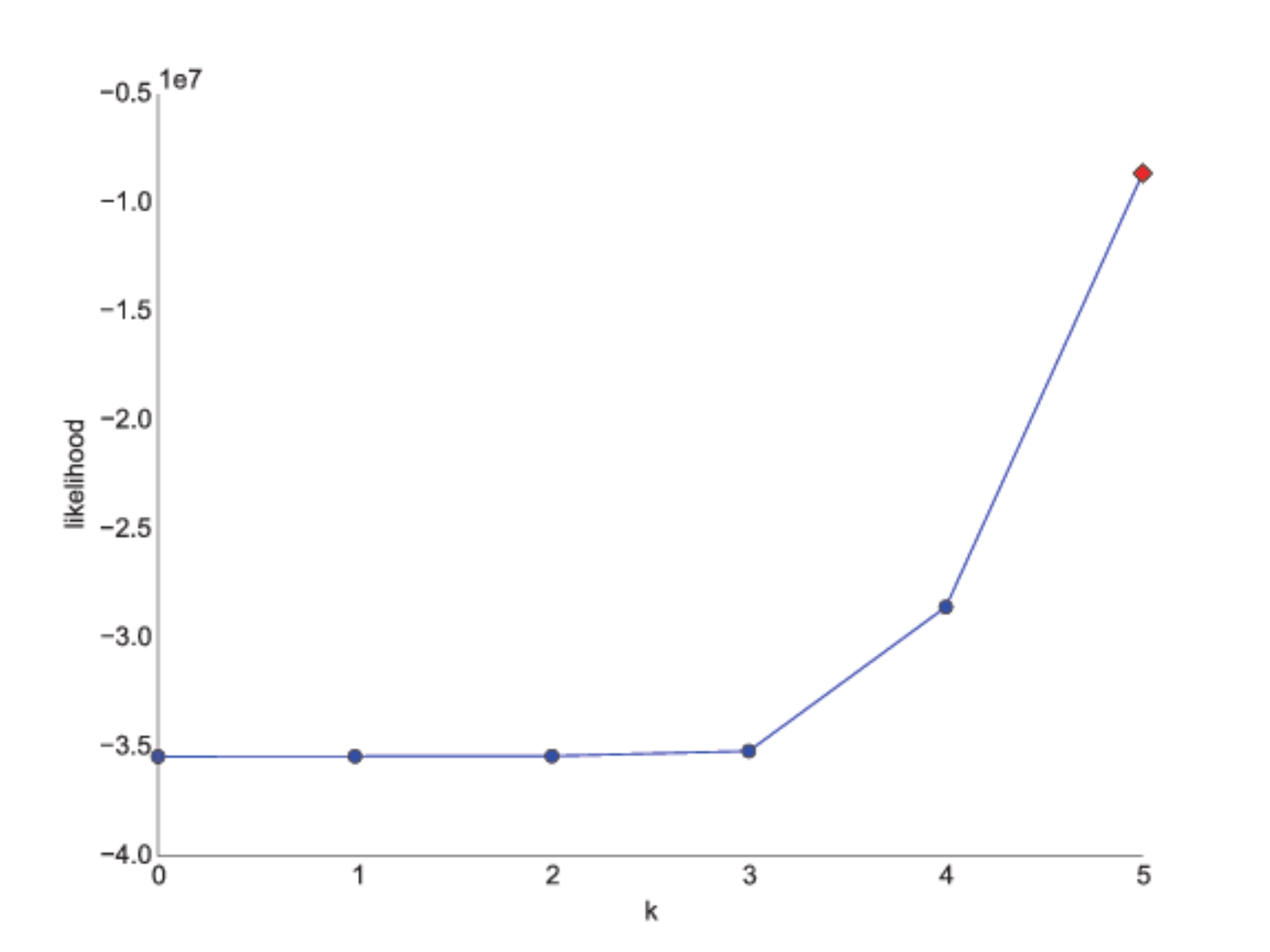}
\caption{\textbf{Log-likelihoods for random path dataset.} Simple log-likelihoods of varying Markov chain orders would suggest higher orders as the higher the order the higher the corresponding log-likelihoods are. This suggests that looking at these log-likelihoods is not enough for finding the appropriate Markov chain order as methods are necessary that balance the goodness-of-fit against the number of model parameters.}
\label{fig:randomloglikelihood}
\end{figure}

Our concrete goal is to determine the appropriate order of a Markov chain.
\green{Using the log-likelihoods of the specific
order models is not enough, as we will always get a better fit to our training
data using higher order Markov chains. The reason for this is that lower order models are nested within higher order models. Also, the number of
parameters increases exponentially with $k$ which may result in overfitting
\cite{murphy} since we can always produce better fits to the data with more model
parameters.} To demonstrate this behavior, we produced a random navigational
dataset by randomly (uniformly) picking a next click state out of a list of
arbitrary states. One of these states determines that a path is finished and a
new one begins. With this process we could generate a random path corpus that is
close to one main dataset of this work (Wikigame topic dataset explained in
the section called~"\nameref{sec:experimental}"). Concretely, we as well chose 26 states and the
same number of total clicks. Purely from our intuition, such a process should produce
navigational patterns with an appropriate Markov chain order of zero
or at maximum one. However, if we look at the
log-likelihoods depicted in Figure~\ref{fig:randomloglikelihood} we can observe
that the higher the order the higher the corresponding log likelihoods are.


This strongly suggests that -- as previously explained -- looking at the log-likelihoods is not enough for
finding the appropriate Markov chain order. Hence, we first resort to a well-known statistical likelihood tool for comparing two models -- the so-called \emph{likelihood ratio test}.


This test is suited for comparing the
fit of two composite hypothesis where one model -- the so-called \emph{null
model} $k$ -- is a special case of the \emph{alternative model} $m$. The test
is based on the log likelihood ratio, which expresses how
much more likely the data is with the alternative model than with the null
model. We follow the notation provided by Tong \cite{tong1975} and denote
the ratio as ${_k}\eta{_m}$:
\begin{equation}
 {_k}\eta{_m} = -2(\mathcal{L(P(D|}{\theta}_k)) -
 \mathcal{L(P(D|\theta}_m)))
 \label{eq:lr}
\end{equation}

To address the overfitting problem we perform a significance test on this ratio.
The significance test recognizes whether a better fit to data comes only from
the increased number of parameters.
The test calculates the p-value of the likelihood ratio distribution. Whenever the null model is 
nested within
the alternative model the likelihood ratio approximately follows a $\chi^2$ distribution with degrees of freedom
specified by $(|S|^m-|S|^k)(|S|-1)$. If the p-value is below a specific significance
level we can reject the null hypothesis and prefer the alternative model
\cite{bartlett}\footnote{\green{Note that this method also utilizes mechanisms usually known from the frequentist school; i.e., hypothesis testing.}}.

\green{
Likelihood ratios and corresponding tests have been shown to be a very understandable approach of specifying evidence \cite{perneger}. They also have the advantage of specifying a clear value (i.e., the likelihood ratio) with can give us intuitive meaning about the advantage of one model over the other.
However, the likelihood-ratio test also has limitations like that it only works for nested models, which is fine for our approach but may be problematic for other use cases. 
It also requires us to use elements from frequentist approaches (i.e., the p-value) for deciding between two models which have been criticized in the past (e.g., \cite{morrison2006significance}).
Furthermore, we are only able to compare two models with each other at a time. This makes it difficult to choose one single model as the most likely one as we may end up with several statistical significant improvements. Also, as we increase the number of hypothesis in our test, we as well increase the probability that we find at least one significant result (Type 1 error)\footnote{\green{We could tackle this problem by e.g., applying the \emph{Bonferroni correction} which we leave open for future work}}. 
}

\subsubsection*{Bayesian Method}
\label{subsubsec:bayesianinference}

\green{
Bayesian inference is a statistical method utilizing the Bayes' rule -- Rev. Thomas Bayes started to talk about the Bayes theorem in 1764 -- for updating prior believes with additional evidence derived from data. A general introduction to Bayesian inference can e.g., be found in \cite{box2011bayesian}; in this article we focus on explaining the application for deriving the appropriate Markov chain order (see \cite{Strelioff} for further details). 
}

In Bayesian inference data and the model parameters are treated as random
variables \green{(cf. MLE where parameters are unknown constants)}. We start with a joint
probability distribution of data $D$ and parameters $\theta_k$ given a model
$M$; that is given a Markov chain of a specified order $k$. Thus, we are
interested in $P(D, \theta_k | M_k)$.

The joint distribution $P(D, \theta_k | M_k)$ can be written as the product of
the conditional probability of data $D$ given the parameters $\theta_k$ and the marginal distribution of the parameters, or
we can write this joint distribution as the product of the conditional probability of the parameters given the data and the marginal distribution of the data.


Solving then for the posterior distribution of parameters given data and a model we obtain the famous Bayes rule:
\begin{equation}
 P(\theta_k | D, M_k) = \frac{P(D | \theta_k, M_k)P(\theta_k | M_k)}{P(D | M_k)},
\end{equation}

 where $P(\theta_k | M_k)$ is the prior probability of model parameters, $P(D | \theta_k, M_k)$ is the likelihood function; that is the probability of observing the data
given the parameters, and $P(D | M_k)$ is the evidence (marginal likelihood).
$P(\theta_k | D, M_k)$ is the posterior probability of the parameters, which we obtain after we update the prior with the data.

For a more detailed and an in-depth technical analysis of Bayesian inference of
Markov chains we point to an excellent discussion of the topic in
\cite{Strelioff}.

\bigskip
\noindent \emph{Likelihood.}
As previously, we have:
\begin{equation}
P(D | \theta_k, M_k) = p(x_1) \prod_i \prod_j p_{ij}^{n_{ij}}
\end{equation}

\bigskip
\noindent \emph{Prior.}
The prior reflects our (subjective or objective) belief about the parameters
before we see the data. In Bayesian inference, conjugate priors are of special
interest.
Conjugate priors result in posterior distributions from the same distribution
family. In our case, each row of the transition matrix follows a categorical
distribution. The conjugate prior for categorical distribution is the Dirichlet
distribution. Further information on applying Dirichlet conjugate prior and
dealing with Dirichlet process can be found in \cite{Huelsenbeck}. The Dirichlet
distribution is defined as $Dir(\alpha)$:
\begin{equation}
 Dir(\alpha) = \frac{\Gamma(\sum_j \alpha_{j})}{\prod_j \Gamma(\alpha_{j})}\prod_j x_{j}^{\alpha_{j} - 1},
\end{equation}

where $\Gamma$ is the gamma function, $\alpha_j > 0$ for each $j$ and $\sum_j x_j = 1$ is a probability simplex. The probability outside of the simplex is $0$.

The \textit{hyperparameters} $\alpha$ reflect our assumptions about the
parameters $\theta$ before we have observed the data. We can think about the
hyperparameters as fake counts in the transition matrix of a Markov
chain. A standard uninformative selection for hyperparameters is a uniform prior
-- for example, we set $\alpha_j = 1$ for each $j$.

Thus, for row $i$ of the transition matrix we have the following prior:
\begin{equation}
Dir(\alpha_{i}) = \frac{\Gamma(\sum_j \alpha_{ij})}{\prod_j \Gamma(\alpha_{ij})}\prod_j p_{ij}^{\alpha_{ij} - 1}
\end{equation}

As before, it holds that:
\begin{equation}
\sum_j p_{ij}=1
\end{equation}

The prior for the complete transition matrix is the product of the Dirichlet distributions for each row:
\begin{equation}
P(\theta_k | M_k) = \prod_i\frac{\Gamma(\sum_j \alpha_{ij})}{\prod_j \Gamma(\alpha_{ij})}\prod_j p_{ij}^{\alpha_{ij} - 1}
\end{equation}


\bigskip
\noindent \emph{Evidence.}
To calculate the evidence we take a weighted average over all possible values of the parameters $\theta_k$.
Thus, we need to integrate out the parameters $\theta_k$.
\begin{equation}
 P(D | M_k) = \int P(D | \theta_k, M_k)P(\theta_k | M_k)d\theta_k
\end{equation}

\begin{eqnarray}
\nonumber P(D | M_k) &=& \int P(D | \theta_k, M_k)P(\theta_k | M_k)d\theta_k\\
\nonumber &=& \int p(x_1) \prod_i \prod_j p_{ij}^{n_{ij}} \prod_i\frac{\Gamma(\sum_j \alpha_{ij})}{\prod_j \Gamma(\alpha_{ij})}\prod_j p_{ij}^{\alpha_{ij} - 1} d\theta_k\\
\nonumber &=&  p(x_1)\prod_i\frac{\Gamma(\sum_j \alpha_{ij})}{\prod_j \Gamma(\alpha_{ij})} \int \prod_j p_{ij}^{n_{ij}} \prod_j p_{ij}^{\alpha_{ij} - 1} d\theta_k\\
\nonumber &=&  p(x_1)\prod_i\frac{\Gamma(\sum_j \alpha_{ij})}{\prod_j \Gamma(\alpha_{ij})} \int \prod_j p_{ij}^{n_{ij}+\alpha_{ij} - 1} d\theta_k
\end{eqnarray}

Please note, that:
\begin{eqnarray}
\nonumber \int \frac{\Gamma(\sum_j \alpha_{j})}{\prod_j \Gamma(\alpha_{j})}\prod_j x_{j}^{\alpha_{j} - 1} dx &=& 1\\
\nonumber \frac{\Gamma(\sum_j \alpha_{j})}{\prod_j \Gamma(\alpha_{j})} \int \prod_j x_{j}^{\alpha_{j} - 1} dx &=& 1\\
\nonumber \int \prod_j x_{j}^{\alpha_{j} - 1} dx &=& \frac{\prod_j \Gamma(\alpha_{j})}{\Gamma(\sum_j \alpha_{j})}
\end{eqnarray}

Thus, we have
\begin{equation}
\int \prod_j p_{ij}^{n_{ij}+\alpha_{ij} - 1} d\theta_k = \frac{\prod_j \Gamma(n_{ij}+\alpha_{ij})}{\Gamma(\sum_j (n_{ij}+\alpha_{ij}))}
\end{equation}

And thus,
\begin{equation}
 P(D | M_k) = p(x_1)\prod_i\frac{\Gamma(\sum_j \alpha_{ij})}{\prod_j \Gamma(\alpha_{ij})} \frac{\prod_j \Gamma(n_{ij}+\alpha_{ij})}{\Gamma(\sum_j (n_{ij}+\alpha_{ij}))}
\label{eq:evidence}
\end{equation}

\bigskip
\noindent \emph{Posterior.}
For the posterior distribution over the parameters $\theta_k$ we obtain:
\begin{eqnarray}
\nonumber P(\theta_k | D, M_k) &=& \prod_i \prod_j p_{ij}^{n_{ij}} \prod_i \prod_j p_{ij}^{\alpha_{ij} - 1} \frac{\Gamma(\sum_j (n_{ij}+\alpha_{ij}))}{\prod_j \Gamma(n_{ij}+\alpha_{ij})}\\
\nonumber  &=& \prod_i \prod_j p_{ij}^{n_{ij} + \alpha_{ij} - 1} \frac{\Gamma(\sum_j (n_{ij}+\alpha_{ij}))}{\prod_j \Gamma(n_{ij}+\alpha_{ij})}
\end{eqnarray}
This equation is the product of the Dirichlet distributions for each row with parameters $n_{j} + \alpha_{j}$:

\begin{equation}
P(\theta_k | D, M_k) = \prod_i Dir(n_i + \alpha_{i})
\end{equation}

The posterior distribution is a combination of our prior belief and the data that we have observed.
In fact, the expectation and the variance of the posterior distribution are:
\begin{equation}
 E[p_{ij}] = \frac{n_{ij} + \alpha_{ij}}{\sum_j (n_{ij} + \alpha_{ij})}
\end{equation}
\begin{equation}
 Var[(p_{ij}] = \frac{(n_{ij} + \alpha_{ij})(\sum_j (n_{ij} + \alpha_{ij}) - (n_{ij} + \alpha_{ij}))}{(\sum_j (n_{ij} + \alpha_{ij}))^2(\sum_j (n_{ij} + \alpha_{ij}) + 1)}
\end{equation}

We can rewrite the expectation as:
\begin{equation}
 E[p_{ij}] = \frac{1}{\sum_j (n_{ij} + \alpha_{ij})}\left(\sum_j
 n_{ij}\frac{n_{ij}}{\sum_j n_{ij}} + \sum_j
 \alpha_{ij}\frac{\alpha_{ij}}{\sum_j \alpha_{ij}}\right)
\end{equation}

Setting $c=\frac{\sum_j n_{ij}}{\sum_j (n_{ij} + \alpha_{ij})}$, we can rewrite the expectation of the posterior distribution as:
\begin{equation}
 E[p_{ij}] = c\frac{n_{ij}}{\sum_j n_{ij}} + (1-c)\frac{\alpha_{ij}}{\sum_j
 \alpha_{ij}}
\end{equation}

Thus, the posterior expectation is a \textit{convex combination} of the MLE and
the prior. When the number of the observation becomes large ($n_{ij} \gg
\alpha_{ij}$) then $c$ tends to $1$, and the posterior expectation tends to the
MLE.

By setting $\alpha_{ij} = 1$ for each $i$ and $j$ we effectively obtain Laplace's prior; that is we apply Laplace smoothing \cite{murphy}.


For model selection we adopt once more the Bayesian inference (again see \cite{Strelioff} for a thorough discussion). We have a set $M$
of models $M_k$ with varying order $k$ and are interested in deciding between several models (c.f. \cite{kass1995bayes}). We are interested in the joint
probability distribution $P(D, M_k)$ of data $D$ and a model $M_k$. We can
write the joint distribution as a product of a conditional probability (of data
given a model, or of a model given the data) and a prior marginal distribution
(of data or a model) and by solving for the posterior distribution of a model
given the data we again obtain the Bayes rule:
\begin{equation}
 P(M_k | D) = \frac{P(D | M_k) P(M_k)}{P(D)},
\end{equation}
where $P(D)$ is the weighted average over all models $M_k$:
\begin{equation}
 P(D) = \sum_k P(D | M_k) P(M_k).
\end{equation}

The likelihood of data $D$ given a model $M_k$ is the evidence $P(D | M_k)$
given by Equation~\ref{eq:evidence}, which is the weighted average over all
possible model parameters $\theta_k$ given the model $M_k$.

Following Strelioff et al. \cite{Strelioff}, we select two priors over the model set $M$ -- a uniform prior and a prior with
an exponential penalty for the higher order models \cite{Strelioff}.
The uniform
prior assigns the identical probability for each model:
\begin{equation}
 P(M_k) = \frac{1}{|M|}.
\end{equation}

With the uniform prior we obtain the following expression for the posterior
probability of a model $M_k$ given the data:
\begin{equation}
 P(M_k | D) = \frac{P(D | M_k)}{\sum_k P(D | M_k)}.
\end{equation}

The prior with the exponential penalty can be defined as:
\begin{equation}
 P(M_k) = \frac{e^{-|S_k|}}{\sum_k e^{-|S_k|}},
\end{equation}
where $|S_k|$ is the number of states of the model $M_k$ and can be calculated as:
\begin{equation}
 |S_k| = |S|^k(|S| - 1),
\end{equation}
with $|S|$ being the number of states of the model of order $1$.

After solving for the posterior distribution for the prior with the exponential penalty we obtain:
\begin{equation}
 P(M_k | D) = \frac{P(D | M_k) e^{-|S_k|}}{\sum_k P(D | M_k) e^{-|S_k|}}.
\end{equation}

The calculations are best implemented with log-evidence and logarithms of the
gamma function to avoid underflow since the numbers are extremely small. To
implement the sum for the normalizing constant in the denominator we apply the
so-called \textit{log-sum-exp trick} \cite{Durbin}.
First, we calculate the
log-evidence: $log P(D | M_k)$ and then calculate the logarithm of the
normalizing constant $log(C)$:
\begin{equation}
 log(C) = log (\sum_k e^{log(P(D | M_k))}).
\end{equation}
A direct calculation of $e^{log(P(D | M_k))}$ results in an underflow, and thus we pull the largest log-evidence $E_{max} = max(log(P(D | M_k))$ out of the sum:
\begin{equation}
 log(C) = E_{max} + log (\sum_k e^{log(P(D | M_k)) - E_{max}}).
\end{equation}


\green{
One downside of using Bayesian model selection is that it is frequently difficult to calculate Bayes factors. Concretely, it is often complicated to calculate the necessary integral analytically and one needs to resort to various alternatives in order to avoid this problem. Nowadays, several such methods exist: e.g., asymptotic approximation or sampling from the posterior (MCMC, Gibbs) \cite{kass1995bayes}. Also, we need to specify prior distributions for the parameters of each model. As elaborated by Kass and Raftery \cite{kass1995bayes}, one approach is to use the BIC (see the next section entitled "\nameref{subsubsec:informationtheoretic}") which gives an appropriate approximation given one specific prior. 

Compared to the likelihood ratio test (see section entitled~\nameref{subsubsec:mle}), the Bayesian model selection technique does not require the models to be nested. The main benefit of Bayesian model selection is that it includes a natural \emph{Occam's razor} -- i.e., a penalty for too much complexity -- which helps us to avoid overfitting \cite{kass1995bayes,mackay1992bayesian,murray2005note,mackay2003information}. The Occam's razor is a principle that advises to prefer simpler theories over more complex ones. Based on this definition there is no need to include extra complexity control as we e.g., additionally did for our exponential penalty. We see this though as a nice further control mechanism for cautiously penalizing model complexity and for validating the natural Occam's razor.
}

\subsubsection*{Information-theoretic Methods}
\label{subsubsec:informationtheoretic}

\green{
Information-theoretic methods are based on concepts and ideas derived from information theory with a specific focus on \emph{entropy}. In the following we will provide a description of the two probably most well-known methods; i.e., AIC and BIC. A thorough overview of information-theoretic methods can e.g., be found in various work by K. P. Burnham~\cite{burnham2002model,burnham2004multimodel}.
}


\smallskip
\noindent \emph{Akaike information criterion (AIC).} 
Akaike \cite{akaike}
introduced in 1973 a one dimensional statistic for determining the optimal model from a class of competing models. The criterion is based on
Kullback-Leibler divergence \cite{KLD} and the asymptotic properties of the
likelihood ratio statistics described in the section entitled "\nameref{subsubsec:mle}". The approach is based on minimization of AIC (minimum AIC estimate -- MAICE)
amongst several competing models \cite{gates1976} and has been first used for
Markov chains by Tong \cite{tong1975}. Hence, we define the AIC based on the
choice of a loss function proposed by Tong \cite{tong1975}:
\begin{equation}
 AIC(k) = {_k}\eta{_m} - 2(|S|^m-|S|^k)(|S|-1)
 \label{eq:aic}
\end{equation}

The test represents an asymptotic version of the likelihood ratio test defined
in Equation~\ref{eq:lr} for composite hypothesis. The idea is to choose $m$
reasonably high and test lower order models until an optimal order is found.
MAICE chooses the order $k$ which exhibits the minimum
AIC score and tries to balance between overfitting and underfitting
\cite{gates1976}.

\smallskip
\noindent \emph{Bayesian Information Criterion (BIC).} In 1978 Schwarz
\cite{schwarz} introduced this criterion which can be seen as an approximation of the Bayes factor for Bayesian model selection (see the previous section entitled "\nameref{subsubsec:bayesianinference}"). It is similar to the AIC introduced
above with the difference that it penalizes higher order models even more by adding an additional penalization for the number of observations \cite{katz}:
\begin{equation}
 BIC(k) = {_k}\eta{_m} - (|S|^m-|S|^k)(|S|-1)ln(n)
 \label{eq:bic}
\end{equation}

Again we choose $m$ reasonably high and test lower order models against it. The
penalty function is the degree of freedom multiplied with the natural logarithm
of the number of observations $n$. This function converges to infinity at a
still slow enough rate and hence, grants a consistent estimator of the Markov
chain order \cite{katz}.


\green{
Frequently, both AIC and BIC suggest the same model. However, there are certain cases, where they might slightly disagree. In model selection literature there is a still ongoing debate of whether one should prefer AIC or BIC over each other -- e.g., see \cite{weakliem1999critique} for a critique of the BIC for model selection. However, as pointed out by Burnham and Anderson \cite{burnham2004multimodel}, each has its strength and weaknesses in distinct domains. The authors emphasize that both can be seen as either frequentist or Bayesian procedures. In case of inequality, Katz \cite{katz} suggests to investigate the patterns further by
simulating observations and investigate distinct sample sizes.
In this paper we instead apply additional model comparison techniques to further
analyze the data.

The performance of AIC and BIC has also been investigated in the terms of determining the appropriate Markov chain order which is the main goal of this article.  R. W. Katz \cite{katz} pointed out that by using AIC there is the possibility of overestimating the true order independent of how large the data is. Hence, he points out that AIC is an inconsistent method. Contrary, he emphasizes that BIC is a consistent estimator -- i.e., if there is a true underlying model BIC will select it with enough data. Alas, it does not perform well for small sample sizes (see also \cite{csiszar2000consistency}). 
Nonetheless, AIC is the most used estimator for determining the appropriate order, maybe due to higher efficiency for smaller data samples, as elaborated by Baigorri et al. \cite{baigorri2009markov}.

While both AIC and BIC seem at first to be very similar to the likelihood ratio test (see section entitled "\nameref{subsubsec:mle}) there are some elementary differences. First and foremost, they can also be applied for non-nested models \cite{burnham2002model}. Moreover, they do not need to resort to hypothesis testing. BIC is also closely related to Bayesian model selection techniques; specifically to the Bayes factor (see section called "\nameref{subsubsec:bayesianinference}"). Kass and Raftery \cite{kass1995bayes} emphasize the advantages of BIC over the Bayes factor by pointing out that it can be applied even when the priors are hard to set. Also, it can be a rough approximation to the logarithm of the Bayes factor if the number of observations is large. BIC is also declared as being well suited for scientific reporting. 

Finally, we want to point out that one could also see AIC as being best for prediction, while BIC might be better for explanation. Also, as pointed out by M. Stone \cite{stone}, AIC is asymptotically equivalent to cross validation (see the section entitled "\nameref{subsubsec:prediction}") if both use maximum likelihood estimation.
}


\subsubsection*{Cross Validation Method}
\label{subsubsec:prediction}

Another -- quite natural -- way of determining the appropriate order of a Markov
chain is cross-validation \cite{chierichetti,murphy}. The
basic idea is to estimate the parameters on a training set and
validate the results on an independent testing set. In order to reduce
variance we perform a stratified 10-fold cross-validation. In difference
to a classic machine learning scenario, we refer to stratified as a way
of keeping approximately the equal amount of observations in each fold. Thus, we keep
approximately 10\% of all clicks in a single fold.

With this method we focus on prediction of the next user click. Markov chains have been already used to
prefetch the next page that the user most probably will visit on the next click.
In the simplest scenario, this prefetched page is the page with the highest transition probability from the current page.
To measure the prediction accuracy we measure the average rank of the actual page in sorted probabilities from the transition matrix.
Thus, we determine the rank of the next page $x_{n+1}$ in the sorted list of
transition probabilities (expectations of the Bayesian posterior) of the current page $x_{n}$ (see the section named~"\nameref{subsec:markovchains}").
We then average the rank over all observations in
the testing set. Hence, we can formally define the average rank $\overline{r(D_f)}$ of a fold
$D_f$ for some arbitrary model $M_k$ the following way:
\begin{equation}
 \overline{r(D_f)} = \frac{\sum_i \sum_j n_{ij} r_{ij}}{\sum_i \sum_j n_{ij}},
\end{equation}

where $n_{ij}$
is the number of transition from state $x_i$ to state $x_j$ in $D_f$ and $r_{ij}$ denotes the rank of $x_j$ in the $i$-th row of the
transition matrix. 

\green{For ranking the states in a row of the matrix, we resort to \emph{modified competition ranking}. This means that if there is a tie between two or more values, we assign the maximum rank of all ties to each corresponding one; i.e., we leave the gaps before a set of ties (e.g., "14445" ranking). By doing so, we assign the worst possible ranks to ties.
One important
implication of this methodology is that we include a natural penalty (a natural Occam's razor) for higher
order Markov chains. The reason for this is that the transition matrices
generally become sparser the higher the order. Hence, we come up with many more
ties and the chance is higher
that we assign higher ranks for observed transitions in the testing data. The most extreme case happens when we do not have any information available for observations in the testing set (which frequently happens for higher orders); then we assign the maximum rank (i.e., the number of states) to all states.  We finally average the ranks over all folds for a given order and suggest the
model with the lowest average rank.\footnote{\green{In order to confirm our findings we also applied an additional way of determining the accuracy which is motivated by a typical evaluation technique known from link predictors \cite{liben}. Concretely, it counts how frequently the true next click is present in the TopK (k=5) states determined by the probabilities of the transition matrix. In case of ties in the TopK elements we randomly draw from the ties. By applying this method to our data we can mirror the evaluation results obtained by using the described and used ranking technique. Note that we do not explicitly report the additional results of this evaluation method throughout the paper.}}


This method requires priors (i.e., fake
counts; see the section named~"\nameref{subsubsec:bayesianinference}") -- otherwise prediction of unseen states is not
possible. It also resorts to the maximum likelihood estimate for calculating the parameters of the models as described in the section entitled "\nameref{subsubsec:mle}". Also, as shown in the previous section called "\nameref{subsubsec:informationtheoretic}" cross validation has asymptotic equivalence to AIC.


One disadvantage of cross validation methods usually is that the results are dependent on how one splits the data. However, by using our stratified k-fold cross validation approach, we counteract this problem as it matters less of how the data is divided. Yet, by doing so we need to rerun the complete evaluation k times, which leads to high computational expenses compared to the other model selection techniques described earlier and we have to manually decide of which k to use. One main advantage of this method is that eventually each observation is used for both training and testing.
}

\begin{table}[b!]
\caption{\bf{Dataset statistics}}
\centering
\begin{tabular}{|l|l|l|l|} \hline
 & \bf{Wikigame} & \bf{Wikispeedia} & \bf{MSNBC} \\ \hline
\#Page Ids & 360,417 & n/a & n/a\\ \hline
\#Topics & 25 & 15 & 17 \\ \hline
\#Paths & 1,799,015 & 43,772 & 624,383 \\ \hline
\#Visited nodes & 10,758,242 & 259,019  & 4,333,359 \\ \hline
\end{tabular}
\label{tab:datasetfacts}
\end{table}

\section*{Materials}
\label{sec:experimental}

In this paper, we perform experiments on three datasets. While the first
two datasets (WikiGame and Wikispeedia) are representatives of goal-oriented
navigation scenarios (where the target node for each navigation sequence is
known beforehand), the third dataset (MSNBC) is representative of free
navigation on the Web (where we have no knowledge about the targets of navigation).


\paragraph{Wikigame dataset}

This dataset is based on the online game
\emph{TheWikiGame}\footnote{\url{http://thewikigame.com/}}. The game platform
offers a multiplayer game, where users navigate from a randomly selected
Wikipedia page (the start page) to another randomly selected Wikipedia page (the
target page). All pairs of start and target pages are connected through
Wikipedia's underlying network.
The users are only allowed to click on Wikipedia links or on the browser back button to reach the target page, but they are not allowed to use search functionality.

In this study, we only considered click paths of length two or more going through
the main article namespace in Wikipedia. Table~\ref{tab:datasetfacts} shows some main characteristics of our Wikigame dataset.

\begin{figure}[t!] \centering
\includegraphics[width=\textwidth]{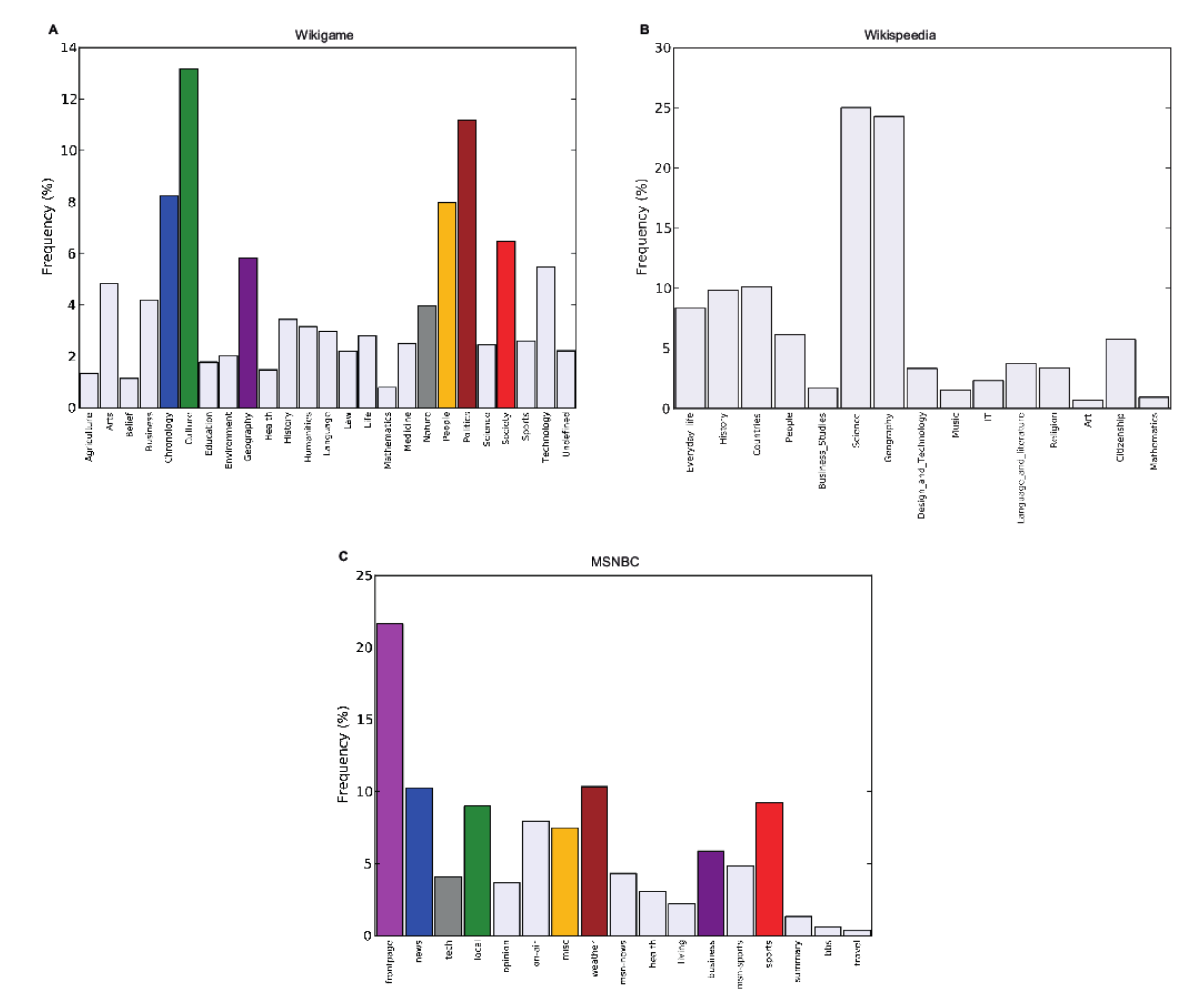}
\caption{\textbf{Topic frequencies.} Frequency of categories (in
percent) of all paths in (A) the Wikigame topic dataset (B) the Wikispeedia
dataset and (C) the MSNBC dataset. The colors indicate the categories we will
investigate in detail later and are representative for a single dataset -- this means that the same
color in the datasets does not represent the same topic.
The Wikigame topic dataset consists of more distinct categories than the
Wikispeedia and MSNBC dataset. Furthermore, the most frequently occuring
topic in the Wikigame topic dataset is Culture with around 13\%.
The Wikispeedia dataset is dominated by the two categories 
the most Science and Geography each making up for almost 25\% of all clicks.
Finally, the most frequent topic in the MSNBC dataset is the frontpage with a
frequency of around 22\%.}
\label{fig:histograms}
\end{figure}

As motivated in Section~"\nameref{sec:intro}", we will represent the navigational
paths through Wikipedia twofold: (a) each node in a path is represented by the
corresponding Wikipedia page ID -- we refer to this as the \emph{Wikigame page}
dataset -- and (b) each node in a path is represented by a corresponding
Wikipedia category (representing a specific topic) -- we call this the \emph{Wikigame topic} dataset. For the
latter dataset we determine a corresponding top level Wikipedia
category\footnote{\url{http://en.wikipedia.org/wiki/Category:Main_topic_classifications}}
in the following way.
The majority of Wikipedia pages belongs to one or more Wikipedia categories. For
each of these categories we find a shortest path to the top level categories and
select a top level category with the shortest distance. In the case of a tie we
pick a top level category uniformly at random. Finally, we replace all
appearances of that page with the chosen top level category. Thus, in this new
dataset we replaced each navigational step over a page with an appropriate
Wikipedia category (topic)  and the dataset contains paths of topics which
users visited during navigation (see Figure \ref{fig:pathexample}). Figure~\ref{fig:histograms} illustrates the distinct topics and their
corresponding occurrence frequency (A).

%
%
%
%
%
%


\paragraph{Wikispeedia dataset}

This dataset is based on a similar online game as the Wikigame dataset called
\emph{Wikispeedia}\footnote{\url{http://www.cs.mcgill.ca/~rwest/wikispeedia/}}.
Again, the players are presented with two randomly chosen Wikipedia pages and
they are as well connected via the underlying link structure of Wikipedia.
Furthermore, users can also select their own start and target page instead of
getting randomly chosen ones. Contrary to the Wikigame, this game is no
multiplayer game and you do not have a time limit. Again, we only look at
navigational paths with at least two nodes in the path.
The main difference to the Wikigame dataset is that Wikispeedia is played on a
limited version of Wikipedia (Wikipedia for
schools\footnote{\url{http://schools-wikipedia.org/}}) with around 4,600 articles.
Some main characteristics are presented in Table~\ref{tab:datasetfacts}.
Conducted research and further explanations of the dataset can be found in
\cite{west,west2,west3,scaria2014last}. 

As we want to look at transitions between topics we determine a
corresponding top level category (topic) for each page in the dataset.
We do this in similar fashion as for our Wikigame dataset, but the Wikipedia version used for
Wikispeedia has distinct top level categories compared to the full Wikipedia.
Figure~\ref{fig:histograms} illustrates the distinct categories and their
corresponding occurrence frequency (B).


\paragraph{MSNBC dataset}

This dataset\footnote{\url{http://kdd.ics.uci.edu/databases/msnbc/msnbc.html}} consists of Web navigational paths from
MSNBC\footnote{\url{http://msnbc.com}} for a complete day. Each single path is a
sequence of page categories visited by a user within a time frame of 24 hours.
The categories are available
through the structure of the site and include categories such as \textit{news}, \textit{tech}, \textit{weather},
\textit{health}, \textit{sports}, etc. In this dataset we also eliminate all paths with just a
single click.
Table~\ref{tab:datasetfacts} shows the basic statistics for this dataset and in
Figure~\ref{fig:histograms} the frequency of all categories of this dataset are
depicted (C).


\paragraph{Data preparation}
Each dataset $D$ consists of a set of paths $\mathbb{P}$. 
A single path contains a single game in the Wikigame and Wikispeedia dataset or
a single navigation session in the MSNBC dataset.
A path $p$ is defined as a $n$-tuple $(v_1,\ldots,v_n)$ with $v_i
\in V, 1\leq i\leq n$ and $(v_i, v_{i+1}) \in E, 1 \leq i \leq n-1$ where $V$ is
the set of all nodes in $\mathbb{P}$ and $E$ is the set of all
observed transitions in $\mathbb{P}$. We also define the
length of a path $len(p)$ as the length of the corresponding tuple
$(v_1,\ldots,v_n)$.
Additionally, we want to define ${\bf p} = \left\{ v_k | k
=1 \ldots n \right\}$ as the set of nodes in a path $p$. Note that $|{\bf
p}| \leq n$.
The finite state set $S$ needed for Markov chain modeling is originally the set
of vertices $V$ in a set of paths $\mathbb{P}$ given a specific dataset $D$.
To prepare the paths for estimation of parameters of a Markov chain of order $k$, we separate single paths by
prepending a sequence of $k$ generic \emph{RESET}
states to each path, and also by appending one \emph{RESET} state at the end of each path.
This enables us to connect independent paths
and -- through the addition of the \emph{RESET} state -- to forget
the history between different paths.
Hence, we end up with an ergodic Markov chain (see \cite{chierichetti}).
With this artificial \emph{RESET} state, the final number of states is $|S|+1$.

\section*{Results}
\label{sec:results}

In this section we present the results obtained from analyzing human
navigation patterns based on our datasets at hand introduced in
Section~"\nameref{sec:experimental}".
We begin by presenting the results of our investigations of memory -- i.e., appropriate Markov chain order using the Markov chain methods thoroughly explained in the section called~"\nameref{sec:methodology}" --
of user navigation patterns in the section entitled~"\nameref{subsec:memory}".
Based on these calculations and observations we
dig deeper into the structure of human navigation and try to find consistent
patterns -- i.e., specific sequences of navigated states -- in
the section named "\nameref{subsec:structure}".

\subsection*{Memory}
\label{subsec:memory}

We start by analyzing human navigation over Wikipedia pages
on the Wikigame page dataset. Afterwards, we will focus on our topic
datasets for getting insights on a topical level.


\subsubsection*{Page navigation}
\label{subsubsec:pagenavi}


\paragraph{Wikigame page dataset}

\begin{figure}[t!]
 \centering
  \includegraphics[width=\textwidth]{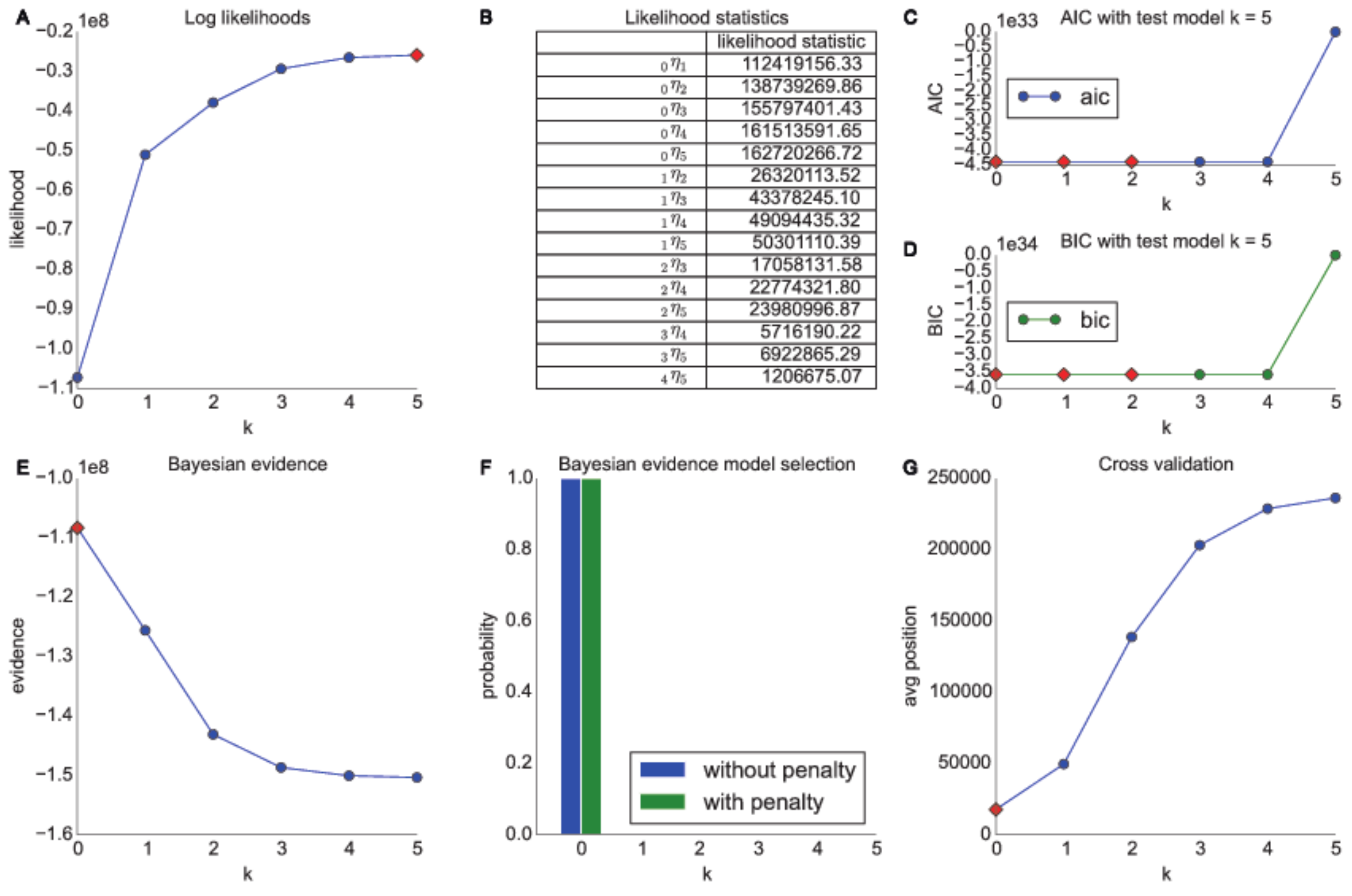}
 \caption{
 \textbf{Model selection results for the Wikigame
 page dataset.}
 The top row shows results obtained using likelihood and information theoretic results: (A)
 likelihoods, (B) likelihood ratio statistics (* statistically significant at
 the 1\% level; ** statistically significant at the 0.1\% level) as well as AIC (C) and
 BIC (D) statistics.
 The bottom row illustrates results obtained from Bayesian Inference: (E) evidence and (F) Bayesian model selection. Finally, the figure
presents the results from (G) cross validation. 
The overall results suggest a zero order Markov chain model.}
 \label{fig:paths_all}
\end{figure}

\green{
The initial Markov chain model selection results (see Figure \ref{fig:paths_all})
obtained from experiments on the Wikigame page dataset confirm our
theoretical considerations. We observe that the likelihoods are rising with
higher Markov chain orders (confirming what \cite{chierichetti} found) which intuitively would indicate a better fit to the data using higher order models. However,
the likelihood grows per definition with increasing order and number of model parameters and
therefore, the likelihood based methods for model selection fail to penalize the
increasing model complexity (c.f. Section~"\nameref{subsubsec:mle}"). All other applied methods take the model complexity
into account. 
}

First, we can imply already from the likelihood
statistics (B) that there might be no
improvement over the most basic zero order Markov chain model as we can not find any statistically significant improvements of higher orders.
Both AIC (C) and BIC (D) results confirm these
observations and also agree with each other. Even though we can see equally
low values for a zero, first and second order Markov chain, we would most likely
prefer the most simple model in such a case -- further following the ideas of the Occam's razor.

In order to extend these primary observations we used a uniform Laplace prior
and Bayesian inference and henceforth, we obtain the results illustrated in the
first two figures of the bottom row in Figure~\ref{fig:paths_all}. The Bayesian inference results again suggest a zero order Markov chain model as the most appropriate as indicated by
the highest evidence (E) and
the highest probability obtained using Bayesian model selection with and without a further exponential penalty for the number of parameters (F).

The observations and preference of using a zero order model are finally confirmed by the
results obtained from using 10-fold cross-validation and a prediction task (G). We can see that the average position is the lowest for a zero order model approving our observations
made above.

\textbf{Summary:} 
\green{
Our analysis of the Wikigame page dataset thereby reveals a
clear trend towards a zero order Markov chain model. This is imminent when looking at all distinct model selection techniques introduced and applied in this article, as they all agree on the choice of weighted random selection as the statistically significant most approvable model.
This is a strong approval of our initial hypothesis stating it is highly difficult to make plausible statements about the appropriate Markov chain order having insufficient data but a vast amount of states. The higher performance of higher order chains can not compensate the necessary additional complexity in terms of statistically significant improvements. However, this may be purely an effect of the data sparsity in our investigation (i.e., the limited number of observations compared to the huge amount of distinct states). One can argue that real human navigation always can be better modeled by at least an order of one, because -- as soon as we have enough data -- links play a vital role in human navigation as humans by definition follow links when they navigate\footnote{Except for teleportation which we do not model in this work.}.
Consequently, we believe that the memoryless Markov chain model is a plausible model for human navigation on a page level. Yet, further detailed studies are necessary to confirm this.
}

At
the same time, one could argue that memory is best studied on a topical level,
where pages are represented by topics. 
Consequently, we focus on studying transitions between topics next, which yields
a reduced state space that allows analysis of the memory and structure of human
navigation patterns on a topical level.


\begin{figure}[t!]
 \centering
 \includegraphics[width=\textwidth]{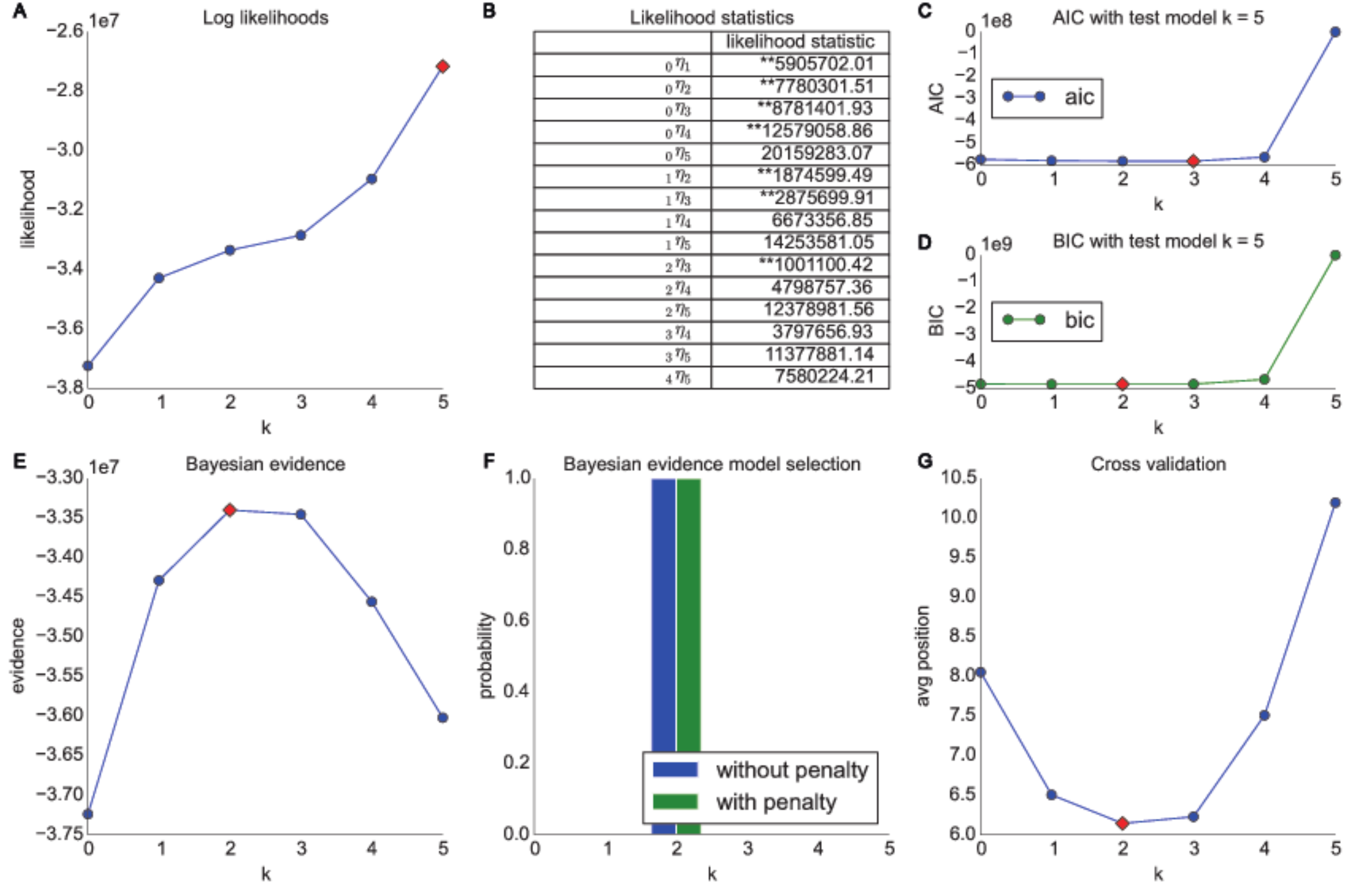}
  \caption{\textbf{
Model selection results for the Wikigame topic dataset.}
 The top row shows results obtained using likelihood and information theoretic results: (A)
 likelihoods, (B) likelihood ratio statistics (* statistically significant at
 the 1\% level; ** statistically significant at the 0.1\% level) as well as AIC (C) and
 BIC (D) statistics. The bottom row illustrates results obtained from Bayesian Inference: (E) shows evidence and (F) Bayesian model selection. (G)
presents the results from cross validation. 
 The overall results suggest that higher order chains seem to be more appropriate for our
 navigation paths consisting of topics. In detail, we find that a second order
 Markov chain model for our Wikigame topic dataset best explains the data.}
 \label{fig:paths_cat}
\end{figure}

\begin{figure}[t!]
 \centering
\includegraphics[width=\textwidth]{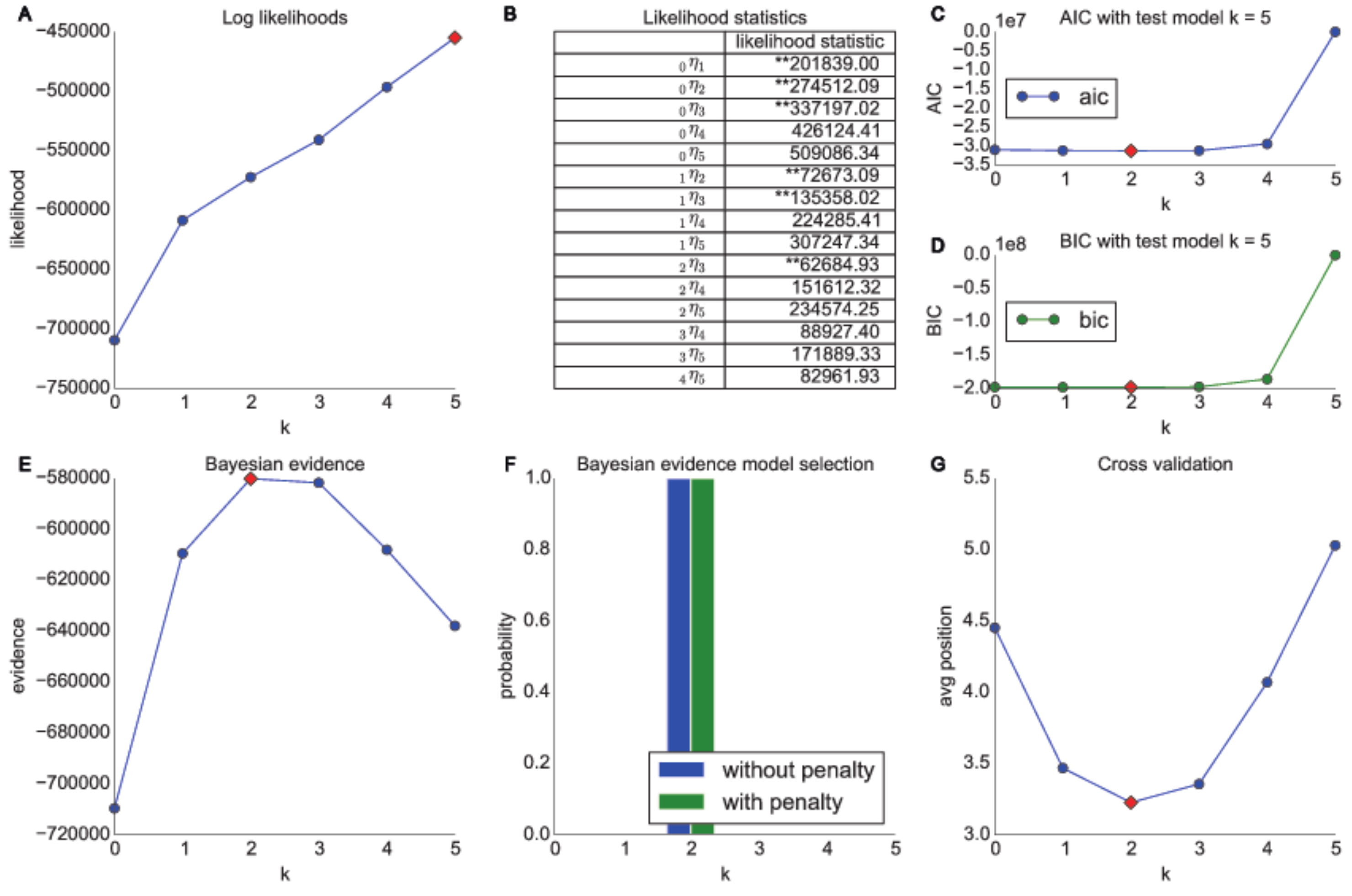}
  \caption{\textbf{
Model selection results for the Wikispeedia dataset.}
 The top row shows results obtained using likelihood and information theoretic results: (A)
 likelihoods, (B) likelihood ratio statistics (* statistically significant at
 the 1\% level; ** statistically significant at the 0.1\% level) as well as AIC (C) and
 BIC (D) statistics. The bottom row illustrates results obtained from Bayesian Inference: (E) shows evidence and (F) Bayesian model selection. (G)
presents the results from cross validation. 
 The overall results suggest that higher order chains seem to be more appropriate for our
 navigation paths consisting of topics. Concretely, we find that a second
 order Markov chain model for our Wikispeedia topic dataset best
 explains the data.}
 \label{fig:paths_wikispeedia}
\end{figure}

\begin{figure}[t!]
 \centering
\includegraphics[width=\textwidth]{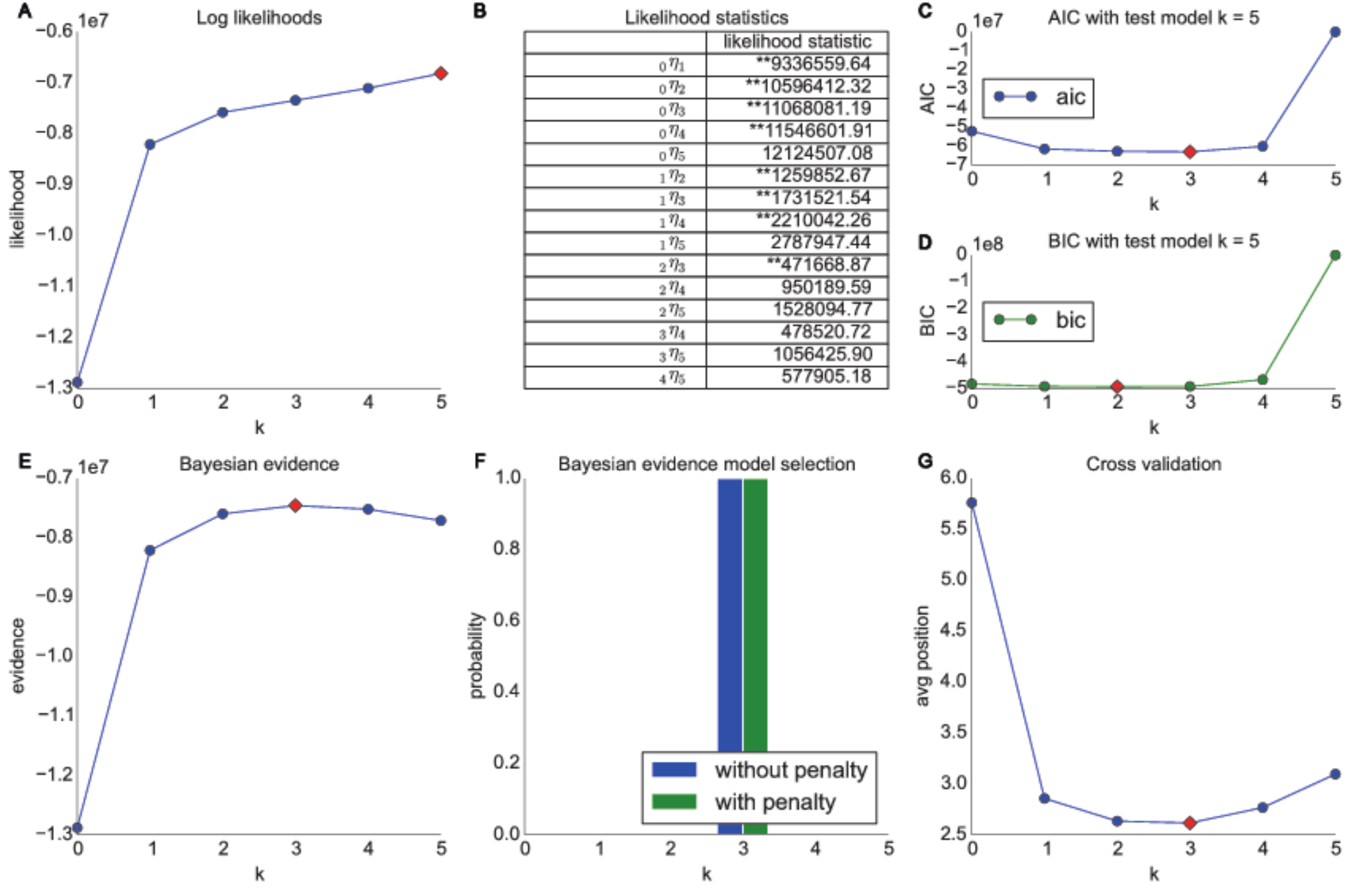}
  \caption{\textbf{
Model selection results for the MSNBC dataset.}
 The top row shows results obtained using likelihood and information theoretic results: (A)
 likelihoods, (B) likelihood ratio statistics (* statistically significant at
 the 1\% level; ** statistically significant at the 0.1\% level) as well as AIC (C) and
 BIC (D) statistics. The bottom row illustrates results obtained from Bayesian Inference: (E) shows evidence and (F) Bayesian model selection. (G)
presents the results from cross validation. 
 The overall results suggest that higher order chains seem to be more appropriate for our
 navigation paths consisting of topics. Specifically, the results suggest a third order Markov
 chain model.}
 \label{fig:paths_msnbc}
\end{figure}

\subsubsection*{Topics navigation}
\label{subsubsec:topicalnavi}

\paragraph{Wikigame topic dataset}

\green{
Performing our analyses by representing Wikipedia pages by their topical
categories shows a much clearer and more interesting picture as one can see in Figure~\ref{fig:paths_cat}.
Similar to above we can see (A) that the log
likelihoods are rising with higher orders. However, in contrast to the Wikigame page
dataset, we can now see (B) that
several higher order Markov chain models are significantly better than lower
orders. In detail, we can see that the appropriate Markov chain order is at
least of order one and we can also observe a trend towards an order of two or
three. Nevertheless, as pointed out in the section entitled "\nameref{subsubsec:mle}", it is hard to concretely suggest one specific Markov chain order from these pairwise comparisons which is why we resort to this extended repertoire of model selection techniques described next.


The AIC (C) and BIC (D) statistics show further
indicators -- even though they are disagreeing -- that the appropriate model is
of higher order. Concretely, the suggest an order of 
three or two respectively by exhibiting the lowest values at these points. 
Not surprisingly, AIC suggests a higher order compared to BIC as the latter model selection method additionally penalized higher orders by the number of observations as stated in the section called "\nameref{subsubsec:informationtheoretic}". 

The Bayesian inference investigations (E, F)
exhibit a clear trend towards a Markov chain of order two. The results in (F) nicely illustrate the inherent Occam's razor of the Bayesian model selection method as both priors -- (a) no penalty and (b) exponential penalty for higher orders -- suggest the same order\footnote{Both priors agree throughout all our investigations in this article.}. 
Finally, the cross validation results (G) confirm
that a second order Markov chain produces the best results, while a third order model is
nearly as good.

\textbf{Summary:} 
Overall, we can see that representing Wikigame paths as
navigational sequences of corresponding topics leads to more interesting
results: Higher order Markov chains exhibit statistically significant
improvements, thereby suggesting that memory effects are at play.
Overall, we can suggest that a second order Markov chain model seems to be the most appropriate for modeling the corresponding data as it gets suggested by all methods except for AIC which is known for slightly overestimating the order.
This means, that humans remember their topical browsing patterns
-- in other words, the next click in navigational
trails is dependent on the previous two clicks on a topical level.
}

\paragraph{Wikispeedia dataset}

This section presents the results obtained from the Wikispeedia dataset
introduced in the section entitled~"\nameref{sec:experimental}". Similar to the Wikigame topic
dataset we look at navigational paths over topical categories in Wikipedia and present the results in Figure~\ref{fig:paths_wikispeedia}.
Again we can observe that the likelihood
statistics suggest higher order Markov chains to be appropriate (B). Yet, further analyses are necessary for a clear choice of the appropriate order.
The AIC (C) and BIC (D) statistics agree
to prefer a second order model; however, we need to note that all orders from zero to four have
similarly low values.
The Bayesian inference investigations (E, F) show a much clearer trend towards a
second order model.
\green{
The prediction results (G) agree on these
observations by also showing the best results for a second order model. This time we can also observe a clear consilience between the cross validation and AIC results which are -- as described in the section called "\nameref{subsubsec:informationtheoretic}" -- asymptotically equivalent.
}

\textbf{Summary:}
This dataset is similar to the Wikigame topic dataset and the results are comparable to the previous results on the first goal-oriented dataset (Wikigame topic).
Hence, even though the game is played on a
much smaller set of Wikipedia articles and also the dataset consists of distinct
categories, we can see the exact same behavior which strongly indicates that
human navigation is not memoryless on a topical level and can be best modeled by a second order Markov chain model.
This strongly suggests that humans follow common topical strategies while navigating
in a goal-oriented scenario.

\paragraph{MSNBC dataset}

In this section we present the results obtained from the MSNBC dataset
introduced in the section called~"\nameref{sec:experimental}". 
Again we look at navigational paths over topical categories and henceforth, we
only look at categorical information of nodes and present the results in Figure~\ref{fig:paths_msnbc}.

Similar to the experiments conducted for the
Wikigame and Wikispeedia topic datasets we can again see, based on the likelihood ratio
statistics (B), that a higher order
Markov chain seems to be appropriate. 
The AIC (C) and BIC (D) statistics suggest an order
of three and two respectively. To further investigate the behavior we illustrate
the Bayesian inference results (E, F) that clearly suggest a third order
Markov chain model. Finally, this is also confirmed by the cross validation
prediction results (G) which again is in accordance with the AIC. 

\textbf{Summary:} By and large, almost all methods for order selection suggest a
Markov chain of order three for the topic sequence in the MSNBC dataset.
Again, we can observe that the navigational patterns are not memoryless. Even though this
dataset is not a goal-oriented navigation dataset, but is based on free
navigation on MSNBC, we can identify similar memory effects as above.


\subsection*{Structure}
\label{subsec:structure}

In the previous section we observed memory patterns in human navigation over topics in information networks. We are now interested in
digging deeper into the structure of human navigational patterns on a topical
level.
Concretely, we are interested in detecting common navigational sequences and in investigating structural
differences between goal-oriented and free form navigation.

First, we want to get a global picture of common transition patterns for
each of the datasets. We start with the Markov chain transition matrices, but instead of normalizing the row vectors,
we normalize each cell by the complete number of transitions in the dataset.
We illustrate these matrices as heatmaps to get insights
into the most common transitions in the complete datasets. Due to tractability,
we focus on a first order analysis and will focus on higher order patterns later
on.

\begin{figure*}[t!]
 \centering
\includegraphics[width=\textwidth]{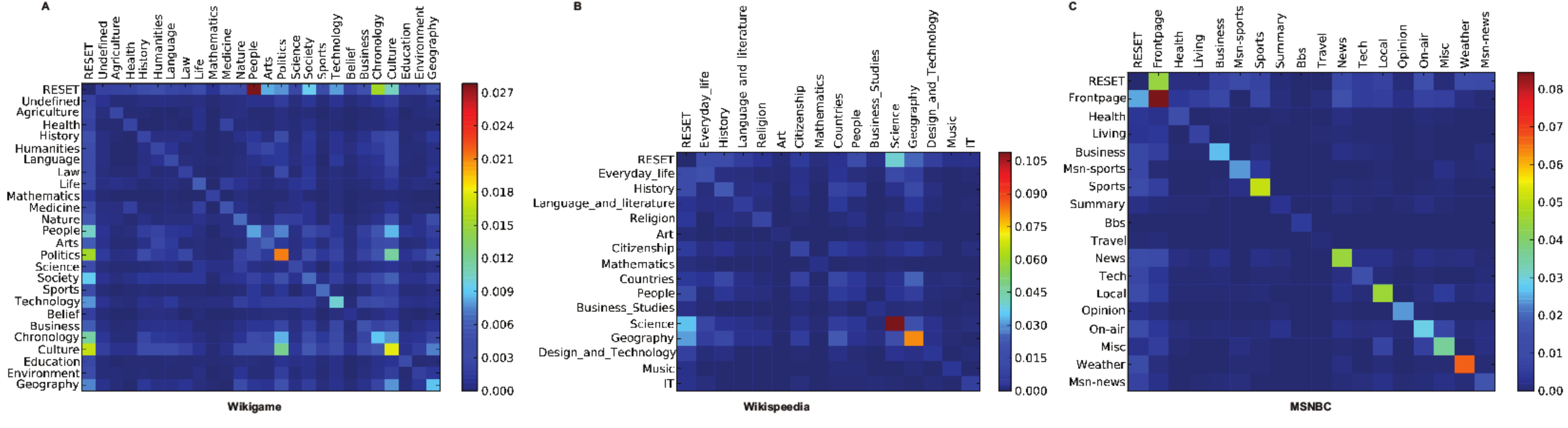}
 \caption{\textbf{Global structure of human navigation.} Common transition patterns of
 navigational behavior on all three topics datasets (Wikigame, Wikispeedia and MSNBC). Patterns are
 illustrated by heatmaps calculated on the first order transition matrices. Each
 cell is normalized by the total number of transitions in the dataset. The
 vertical lines depict starting states and the horicontal lines depict target
 states. A main observation is that self transitions -- e.g., a transition from
 \emph{Culture} to \emph{Culture} -- are dominating all datasets. However, the goal-oriented datasets (Wikigame and Wikispeedia) exhibit more transitions between distinct
 categories than the free navigation dataset (MSNBC).}
 \label{fig:heatmaps}
\end{figure*}

The heatmaps are illustrated in Figure~\ref{fig:heatmaps}. Predominantly, we can
observe that self transitions seem to be very common as we can see from the high
transition counts in the diagonals of the matrices.
This means, that users regularly seem to stay in the same topic while they
navigate the Web\footnote{\green{Consequently, we might get better representations of the data by using Markov chain models that, instead modeling state transitions in equal time steps, additionally stochastically model the duration times in states (e.g., semi Markov or Markov renewal models). However, we leave these investigations open for future work.}}.
For the Wikigame (A) we can observe that the
categories \emph{Culture} and \emph{Politics} are the most visited topics throughout the navigational paths.
Most
of the time the navigational paths start with a page belonging to the
\emph{People} topic which is visible by the dark red cell from \emph{RESET}
to \emph{People} (remember that the \emph{RESET} state marks both the start
and end of a path - see Section~"\nameref{sec:experimental}").
However, as this is a
game-based goal-oriented navigation scenario, the start node is always
predefined.
In our second goal-oriented navigation dataset (B) we can see that the paths are dominated by
transitions from and to the categories \emph{Science} and \emph{Geography} and
there are fewer transitions between other topics. In our MSNBC
dataset (C) we can observe that most of the time users remain in the same topic while they navigate and
globally no topic changes are dominant. This may be an artifact of the free
navigation users practice on MSNBC. Perhaps unsurprisingly, users start with the frontpage most of the time while navigating but do not necessarily come back
to it in the end.

\begin{figure}[t!]
  \centering
\includegraphics[width=\columnwidth]{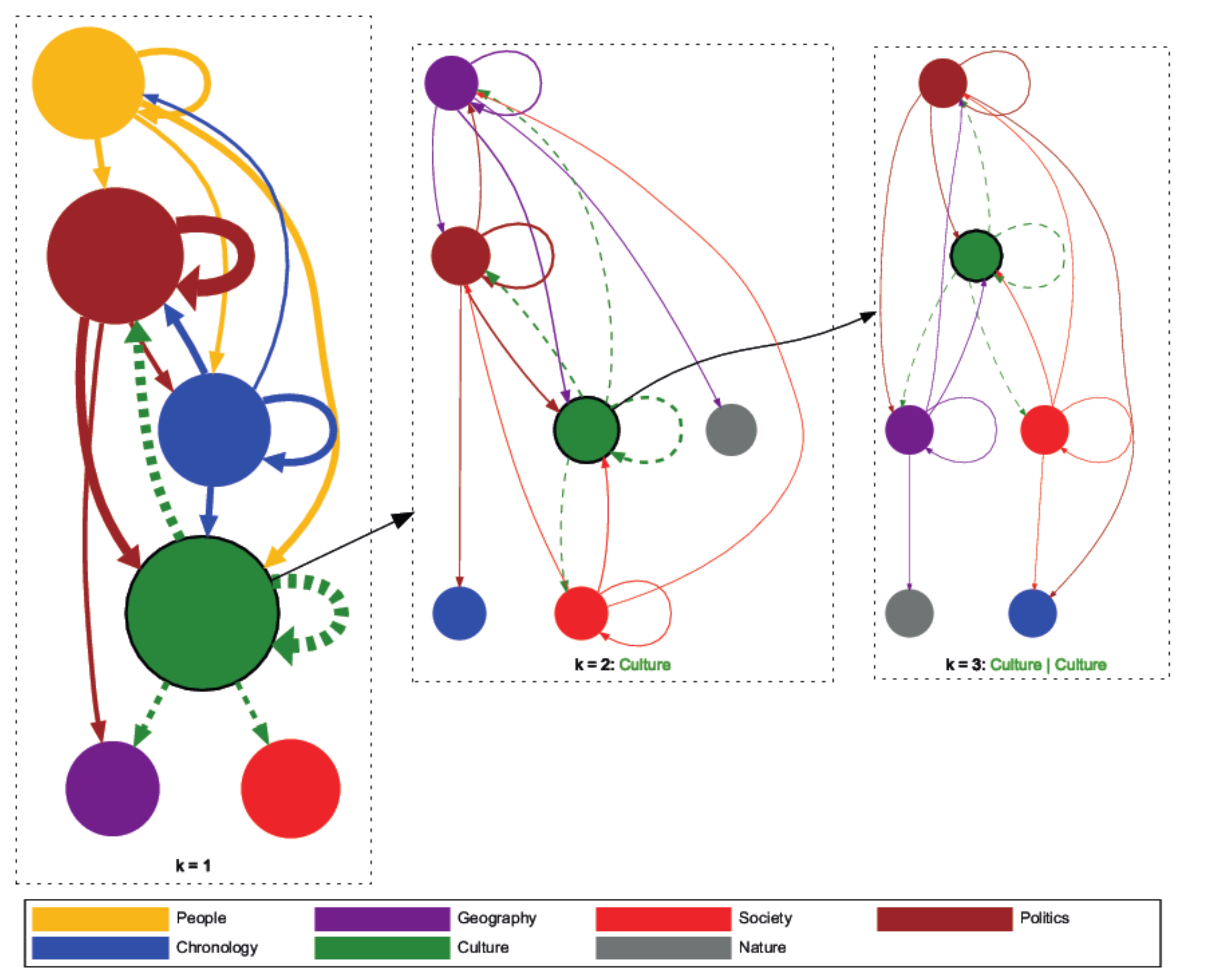}
 \caption{\textbf{Local structure of navigation for the Wikigame topic dataset.} The graphs above illustrate
 selected state transitions from the Wikigame topic dataset for different $k$ values.
 The nodes represent categories and the
links illustrate transitions between categories.
The link weight corresponds to the transition probability from the source to the
target node determined by MLE. The node size corresponds to the sum of the incoming transition probabilities from all other nodes to that source node.
 In the left figure the top four categories with the highest
 incoming transition probabilities are illustrated for an order of $k=1$. For
 those nodes we draw the four highest outgoing transition probabilities to other
 nodes. In the middle figure we visualize the Markov chain of order $k=2$ by
 setting the top topic (\emph{Culture}) as the first click; this diagram
 shows transition probabilities from top four categories given that users first
 visited the \emph{Culture} topic. For example, the links from the red node
 (\emph{Society}) in the bottom-right part of the diagram represent the
 transition probabilities from the sequence (\emph{Culture}, \emph{Society}).
 Similarly, we visualize order $k=3$ in the right figure by selecting a node with the highest
 incoming probability (\emph{Culture}, \emph{Culture}) of order $k=2$. We then
 show transition probabilities from other nodes given that users already visited
 (\emph{Culture}, \emph{Culture}).
 For example, the links from the brown node (\emph{Politics}) at the top
 represent the transition probabilities from the sequence
 (\emph{Culture}, \emph{Culture}, \emph{Politics}).}
\label{fig:local_WIKI_all}
 \end{figure}

 \begin{figure}[t!]
  \centering
\includegraphics[width=\columnwidth]{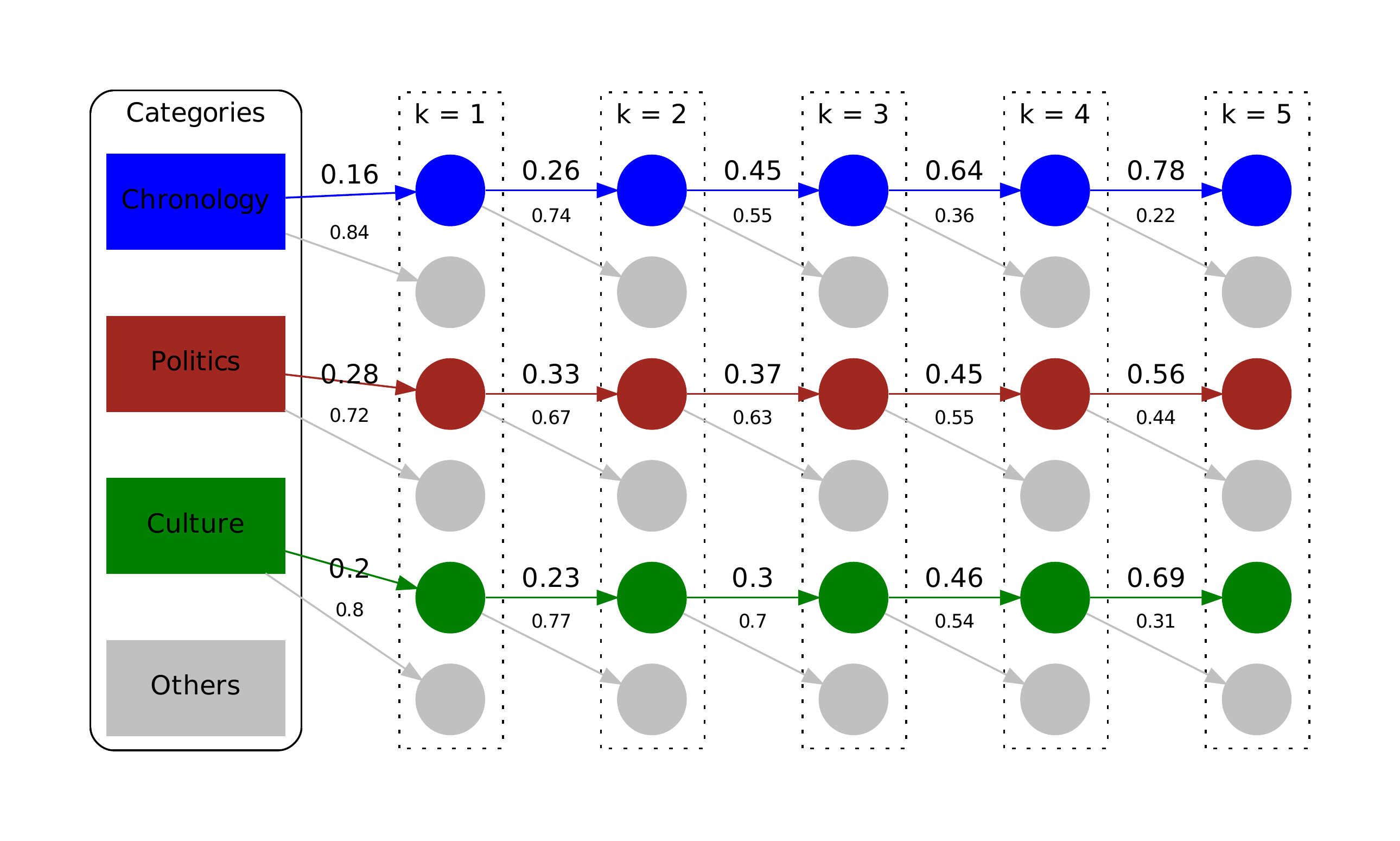}
 \caption{\textbf{Self transition structure of navigation for the Wikigame topic dataset.} The number of times
 users stay within the same topic vs. the number of times they change
 the topic during navigation for different orders $k$ for our Wikigame
 dataset.
 Only the top three categories with the highest transition probabilities are shown.
 With high consistency, the transition probabilities to the same topic increase while those to
 other categories decrease with ascending order $k$.}
\label{fig:sameornot_WIKI}
 \end{figure}

As we have now identified global navigational patterns on the first order
transition matrices we turn our attention to models of higher order. Furthermore, we are
now interested in investigating local transition probabilities -- e.g., being at
topic \emph{Science}, what are the transition probabilities to other states.
The transition weights directly correspond to the transition probabilities from
the source to the target state determined by the MLE (see
the section called~"\nameref{subsubsec:mle}"). We illustrate these local transitional patterns
for our Wikigame dataset in Figure~\ref{fig:local_WIKI_all} (the
investigations on the other goal-oriented Wikispeedia dataset exhibit similar patterns,
but are omitted due to space limitations).
Similar to the observations in Figure~\ref{fig:heatmaps} we can observe that
\emph{Culture} is the most visited topic in our Wikigame dataset. We can now
also identify specific prominent topical transition trails. For example, users
seem to navigate between \emph{Culture} and \emph{Politics} quite frequently and
also vice versa. Contrary, there seem to be specific unidirectional patterns too, e.g., users frequently navigate from \emph{People} to
\emph{Politics} but not vice versa. Higher order chains also show similar
structure, but on a more detailed level.
As previously, the figure also depicts that the vast amount of transitions is
between same categories. However, we can now observe that this is also the case
for higher order Markov chains -- this suggests, that the probability that users stay in the
same topic increases with each new click on that topic.

To further look into this structural pattern, we illustrate the number of times
users stay within the same topic vs. the number of times they change the
topic during navigation in Figure~\ref{fig:sameornot_WIKI}. We can see that
the longer the history -- i.e., the higher the order of the Markov chain -- the
more likely people tend to stay in the same topic instead of switching to
another topic. We can also see differences
regarding this behavior between distinct categories; e.g., users are more likely
to stay in the topic \emph{Chronology} than in the topic \emph{Politics}
the higher the order is.
For our Wikispeedia dataset we can observe similar patterns --
i.e., the higher the order the higher the chance to stay in the same topic.

\begin{figure}[t!]
  \centering
\includegraphics[width=\columnwidth]{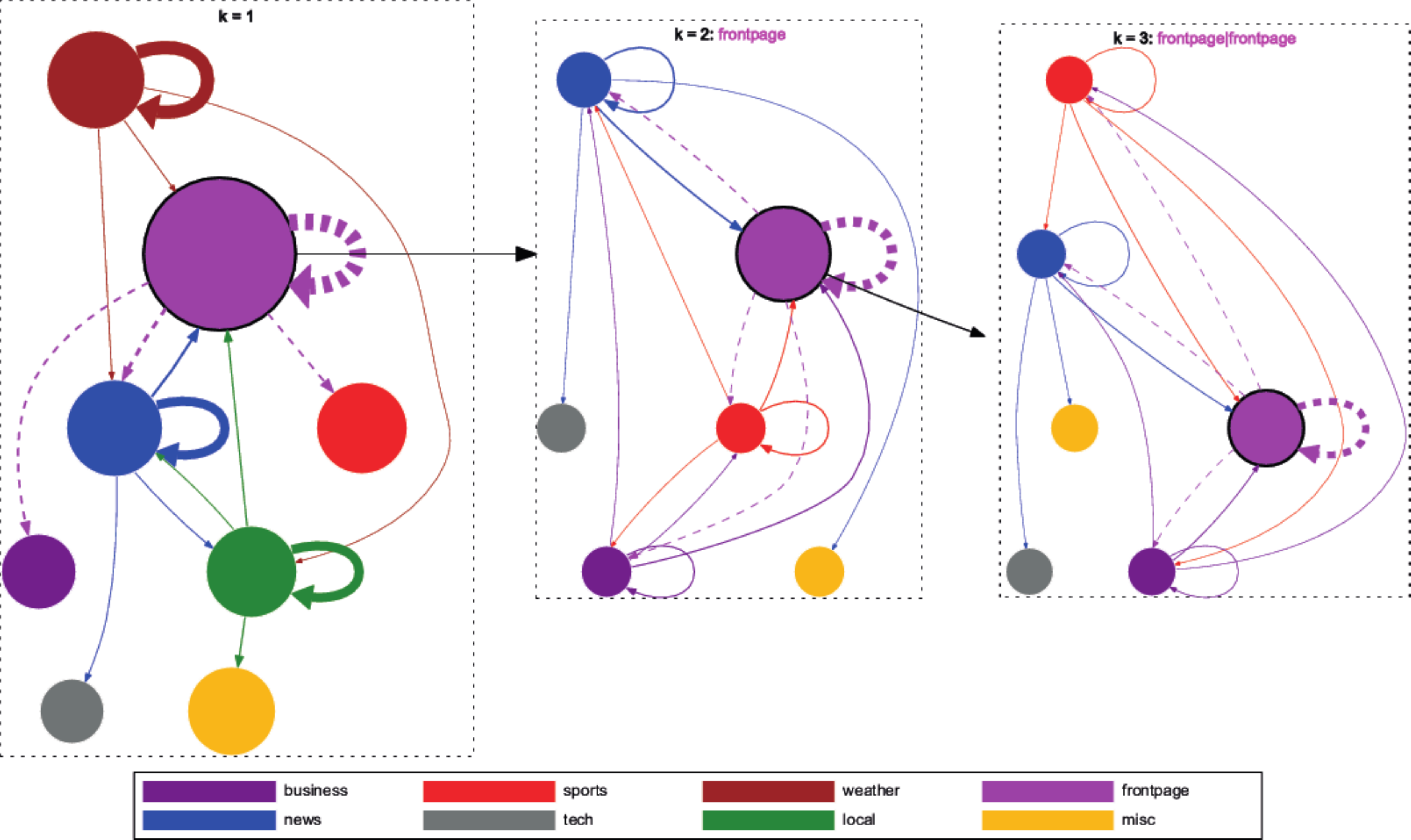}
 \caption{\textbf{Local structure of navigation for the MSNBC dataset.}  The
 graphs above illustrate selected state transitions from the MSNBC dataset for different $k$ values.
 The nodes represent categories and the
links illustrate transitions between categories.
The link weight corresponds to the transition probability from the source to the
target node determined by MLE. 
The node size represents the global importance of a node in the whole dataset 
and corresponds to the sum of the outgoing transition
probabilities from that node to all other nodes.
For visualization reasons we primarily focus on the top four categories with the
highest sum of outgoing transition probabilities -- i.e., those with the largest
node sizes -- for an order of $k=1$. 
For those nodes we draw the four highest outgoing transition probabilities to
other nodes.
In the middle figure we visualize the Markov chain of order $k=2$ by setting the
top topic (frontpage) from order $k=1$ as the first click; this diagram shows transition
probabilities from top four categories given that users first visited
the frontpage topic (represented by the dashed transitions in the left
figure representing $k=1$).
For example, the links from the blue node (news) in the top-left corner of the diagram represent the transition probabilities from
the sequence (frontpage, news) to other nodes.
Similarly, we visualize order $k=3$ in the right figure by selecting a node with the highest sum of outgoing transition probabilities (frontpage, frontpage) and its four highest outgoing transition probabilities from order $k=2$ (represented by the dashed transitions in the middle figure representing $k=2$). We then show transition probabilities
from other nodes given that users already visited (frontpage, frontpage). For
example, the links from the red node (sports) at the top represent the transition
probabilities from the sequence (frontpage, frontpage, sports) to other nodes.}
\label{fig:local_msnbc_all}
 \end{figure}

\begin{figure}[t!]
  \centering
\includegraphics[width=\columnwidth]{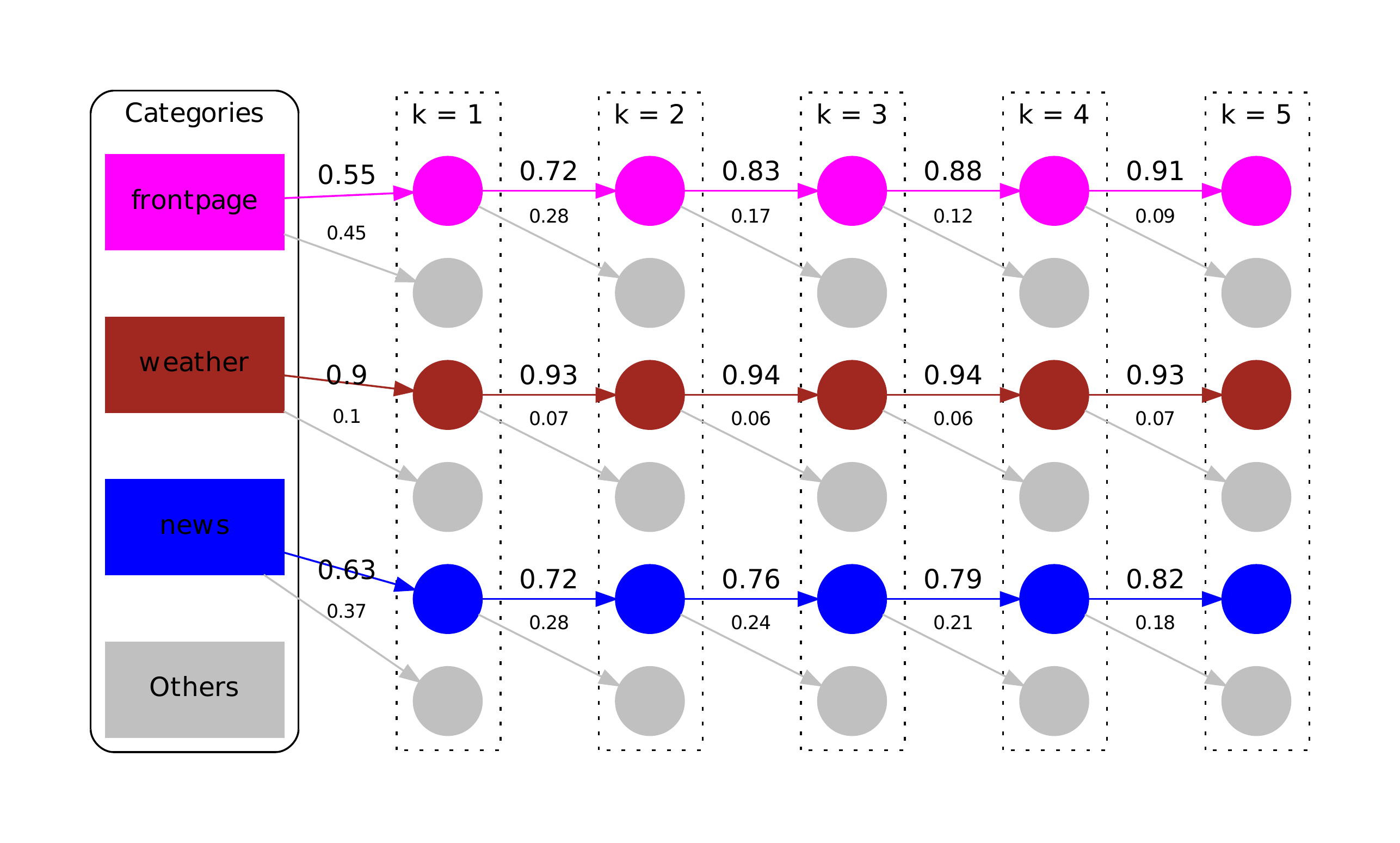}
 \caption{\textbf{Self transition structure of navigation for the MSNBC dataset.} The
 number of times users stay within the same topic vs. the number of times they change the topic during navigation for different 
 values of $k$. Only the top three categories with the highest transition probabilities are shown. 
 With high consistency, the transition probabilities to the same topic increase while those to 
 other categories decrease with ascending order $k$.}
\label{fig:sameornot_msnbc}
 \end{figure}

 In order to contrast goal-oriented and free-form navigation, we also depict state transitions in
 similar fashion derived from the MSNBC dataset in
 Figure~\ref{fig:local_msnbc_all}. In this figure we can see that the topic
 \emph{business} is the most used. To give a navigational example: users
 frequently navigate from \emph{business} to \emph{news} and vice versa.
 However, there are also navigational patterns just going one direction. For
 example, users seem to frequently navigate from \emph{business} to
 \emph{sports} but not in the opposite direction. Again, higher order chains show similar patterns.
 Like in the Wikigame topic dataset we can as well observe that most of the
 transitions seem to be between similar categories. In Figure~
 \ref{fig:sameornot_msnbc} we depict the number of times a user stays in the
 same topic vs. the number of times she switches the topic for the
 categories with the highest transition probabilities. We can again observe that
 the higher the Markov chain the more likely people tend to stay in the same
 topic while navigating. Nevertheless, an interesting difference to the
 Wikigame topic dataset can be observed. Concretely, we can see that the
 probability of staying in the same topic is much higher for the MSNBC
 dataset. Especially, the topic \emph{weather} exhibits a very high
 probability of staying in the same topic ($0.9$ for $k=1$). A possible
 explanation is that users navigate on a semantically more narrow path on MSNBC.
 If you are interested about the weather you just check the specific pages on
 MSNBC while on Wikipedia you might get distracted by different categories at a
 higher probability. So these concrete observations seem to be very specific for
 the Web site and domains of the site users navigate on while the general
 patterns seem to be applicable for both of our datasets at hand.


\subsection*{Discussion}
\label{subsec:discussion}
Our findings and observations in this article show
that simple likelihood investigations (see e.g., \cite{chierichetti}) may not be
sufficient to select the appropriate order of Markov chains and to prove or
falsify whether human navigation is memoryless or not.
To ultimately answer this, we think it is inevitable to look deeper into the
results obtained and to investigate them with a broader spectrum of model selection
methods starting with the ones presented in this work.

By applying these methods to human navigational data, the results suggest that on the Wikigame
page dataset a zero order model should be preferred.
This is due to the rising complexity of higher order models and indicates that
it is difficult to derive the appropriate order for finite datasets with a huge
amount of distinct pages having only limited observations of human navigational
behavior. 
In this article we presented and applied a variety of distinct 
model selection that all include (necessary)
ways of penalizing the large number of parameters needed for higher order models.
\green{Yet, we do not necessarily know what would happen if we would apply the models to a much larger number of navigational paths over pages. Perhaps higher order models would then outperform lower ones. As it is unlikely to get hands on such an amount of data for large websites, a starting point to further test this could be to analyze a sub-domain with rich data; i.e., a large number of observations over just a very limited number of distinct pages. However, due to no current access to such data, we leave this open for future work.}


On the other hand, the results on a topical level are intriguing and show a
much clearer picture: They suggest that the navigational patterns are not memoryless. Higher order
Markov chains -- i.e., second or third order -- seem to be the most appropriate.
Henceforth, the navigation history of users seem to span at least two or three
states on a topical level. This gives high indications that common strategies
(at least on a topical level) exist among users navigating information
networks on the Web. It is certainly intriguing to see similar memory patterns in both
goal-oriented navigation (Wikigame and Wikispeedia) and free form navigation
(MSNBC), and different kinds of systems (encylopedia vs. news portal).

In order
to confirm that these observed memory effects are based on the actual human
navigation patterns we again look at our random path dataset introduced in
the section entitled~"\nameref{subsubsec:mle}" with the log-likelihoods visualized in
Figure~\ref{fig:randomloglikelihood}. We can recapitalize, that these simple
log-likelihoods would suggest a higher order model for the randomly produced
navigational patterns. However, if we apply our various model selection
techniques the results suggest a zero or at maximum a first order Markov chain
model which is the logic conclusion for this random process. Hence, this
confirms that our observations on the real nature navigational data are based on
human navigational memory patterns and would not be present in a random
process.

\begin{figure}[t!]
 \centering
\includegraphics[width=\textwidth]{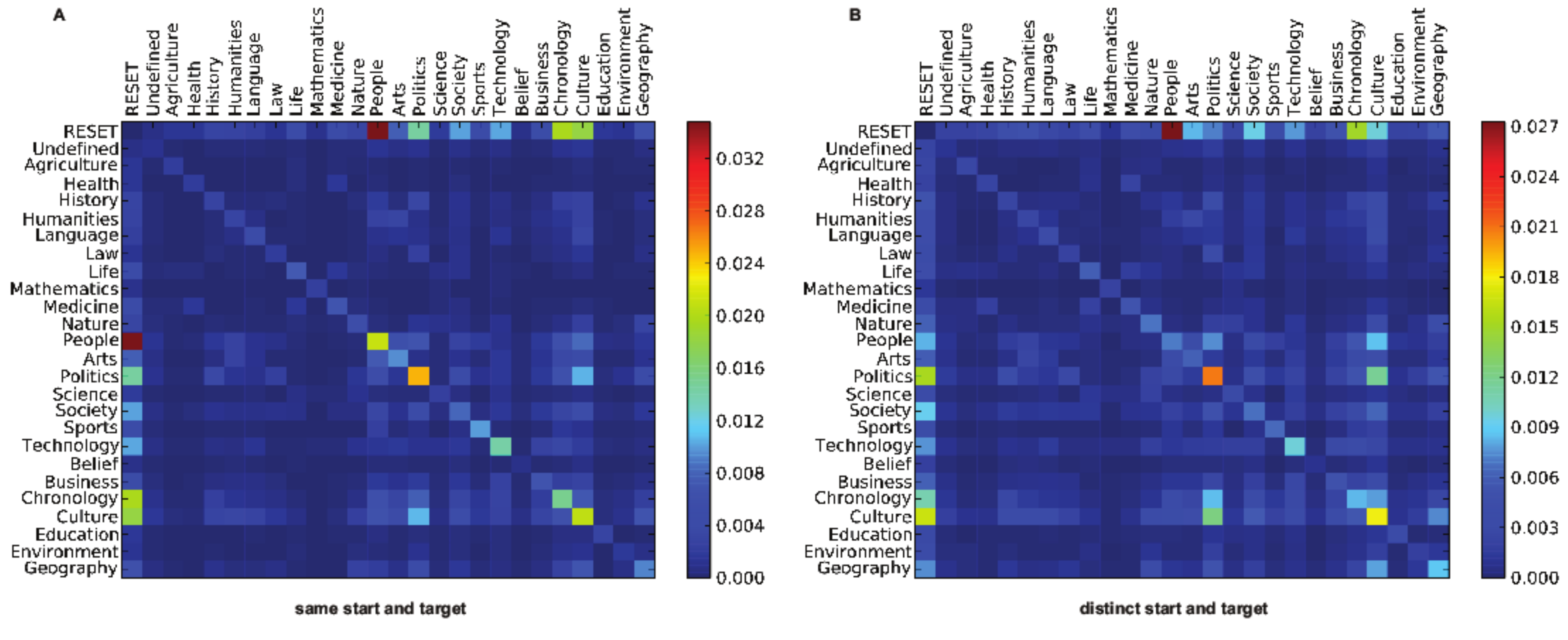}
 \caption{\textbf{Common global transition patterns of navigational behavior on the
 Wikigame topic dataset.} The results should be compare with Figure~\ref{fig:heatmaps}. The results
 are split by only looking at a corpus of paths where each path starts with the same topic as it ends (A) and by looking at a corpus with
 distinct start and target categories (B).
}
 \label{fig:heatmaps_same_diff}
\end{figure}

Finally, we showed in the section called~"\nameref{subsec:structure}" that common structure in
the navigational trails exist among many users -- i.e., common sequences of
navigational transitions. First of all, we could observe that transitions
between the same topic are common among all three datasets. However,
they occur more frequently in our free form navigational data (MSNBC) than in
the goal-oriented navigation datasets (Wikigame and Wikispeedia).
Furthermore, users also seem to be more likely to stay longer in the same topic while
navigating MSNBC while they seem to switch categories more frequently in both
the Wikigame and Wikispeedia datasets. A possible explanation for this user
behavior might be that users on MSNBC are more driven by specific information
needs regarding one topic. For example, a user might visit the website to get
information about the weather only. Contrary, exact information goals on
Wikipedia might not always be in the same topic. Suppose, you are
located on \emph{Seoul} which belongs to the \emph{Geography} topic and you
want to know more about important inventions made in \emph{Seoul}. A possible
path then could be that you navigate over a \emph{People} topic page and
finally reach a \emph{Science} topic page.
However, we need to keep in mind that our goal-oriented datasets are based on
game data with predefined start and target nodes. This means, that if the target nodes
regularly lie in distinct categories, the user might
be forced to switch categories more frequently. To rule this out, we illustrate
the heatmap of our Wikigame dataset (cf. Figure~\ref{fig:heatmaps}) again by splitting
the path corpus into two parts (see Figure~\ref{fig:heatmaps_same_diff}): (A) only considering  paths where the start
and target node lie in the same topic  and (B) only taking paths with distinct
start and target categories.
If the bias of given start and target nodes would influence our observations for
specific structural properties of goal-oriented navigational patterns,
Figure~\ref{fig:heatmaps_same_diff} would show strong dissimilarities between
both illustrations which is not the case. Hence, we can state with strong confidence that the
differences between goal-oriented and free form navigation stated in this section are truly
based on the distinct strategies and navigational scenarios.
Nevertheless, we also need to keep
in mind that the website design and inherent link structure (Wikipedia vs. MSNBC) might also influence this
behavior. For example, a reason could be that Wikipedia has more direct
links between distinct categories in comparison to MSNBC or that Wikipedia's historical coverage steers user behavior to specific kinds of navigational patterns. To explicitly rule
this possibility out, we would need to investigate the underlying link networks
in greater detail, which we leave open for future work. We also plan on looking at data capturing navigational paths over distinct platforms of the Web (e.g., from toolbar data) which may allow us to 
make even more generic statements about human navigation on the Web.

\section*{Conclusions}
\label{sec:conclusions}

This work presented an extensive view on detecting memory and structure in
human navigational patterns. We leveraged Markov chain models of varying order
for detecting memory of human navigation and took a thorough look at
structural properties of human navigation by investigating Markov chain
transition matrices.

We developed an open source framework\footnote{\url{https://github.com/psinger/PathTools}} \cite{github} for detecting memory of human navigational patterns
by calculating the appropriate Markov chain order using four different, yet
complementary, approaches (likelihood, Bayesian, information-theoretic and
cross validation methods). In this article we thoroughly present each method and emphasize strengths, weaknesses and relations between them.  By applying this framework to actual human navigational data we find that it is indeed
difficult to make plausible statements about the appropriate order of a Markov
chain having insufficient data but a vast amount of states which results in too complex models.
However, by representing pages by their corresponding topic
we could identify that navigation on a topical level is not memoryless -- an
order of two and respectively three best explain the observed data,
independent whether the navigation is goal-oriented or free-form. Finally, our
structural investigations illustrated that users tend to stay in the same topic while
navigating. However, this is much more frequent for our free form navigational
dataset (MSNBC) as compared to both of the goal-oriented datasets (Wikigame and Wikispeedia).


Future attempts of modeling human behavior in the Web can benefit from the
methodological framework presented in this work to thoroughly investigate such
behavior. If one wants to resort to a single model selection technique, we would recommend to use the Bayesian approach if computationally feasible. 

Our work strongly indicates memory effects of human
navigational patterns on a topical level. Such observations as well as
detailed insights into structural regularities in human navigation patterns can
e.g., be useful for improving recommendation systems, web site design as well as
faceted browsing.
In future work, we want to extend our ideas of representing Web pages with
categories by looking at further features for representation. We also
plan on tapping into the usefulness of further Markov models like the hidden Markov model, varying order Markov model or semi Markov model. Also, we want to improve recommendation algorithms by the insights generated in this work
and explore the implications higher order Markov chain models may have on
ranking algorithms like PageRank.

\section*{Acknowledgments}
We want to thank Alex Clemesha (Wikigame) and Robert
West (Wikispeedia) for granting us access to their navigational datasets as well as the anonymous reviewers for the highly valuable inputs.

\bibliography{sample}

\end{document}